\documentclass[sigconf]{acmart}
\usepackage{framed}   
\usepackage{float}
\usepackage{stfloats}
\usepackage{listings}

\newenvironment{graybox}[1]{%
  \par\vspace{0.5em}
  \noindent\colorbox{gray!75!black}{%
    \parbox[t]{\dimexpr\linewidth-2\fboxsep}{%
      \raggedright\textcolor{white}{\textbf{#1}}%
    }%
  }\par\nointerlineskip\noindent
  \MakeFramed{\advance\hsize-\width\FrameRestore}%
  \raggedright\noindent
}{%
  \endMakeFramed
}

\AtBeginDocument{%
  }

\copyrightyear{2026}
\acmYear{2026}
\setcopyright{cc}
\setcctype{by}
\acmConference[CHI '26]{Proceedings of the 2026 CHI Conference on Human Factors in Computing Systems}{April 13--17, 2026}{Barcelona, Spain}
\acmBooktitle{Proceedings of the 2026 CHI Conference on Human Factors in Computing Systems (CHI '26), April 13--17, 2026, Barcelona, Spain}
\acmPrice{}
\acmDOI{10.1145/3772318.3790910}
\acmISBN{979-8-4007-2278-3/2026/04}

\begin{document}

\title[Navigating Uncertainty]{Navigating Uncertainties: How GenAI Developers Document Their Models on Open-Source Platforms}

\author{Ningjing Tang}
\affiliation{%
  \institution{Carnegie Mellon University}
  \city{Pittsburgh}
  \state{Pennsylvania}
  \country{USA}
}
\email{ningjint@andrew.cmu.edu}

\author{Megan Li}
\affiliation{%
  \institution{Carnegie Mellon University}
  \city{Pittsburgh}
  \state{Pennsylvania}
  \country{USA}
}
\email{meganli@andrew.cmu.edu}

\author{Amy Winecoff}
\affiliation{%
  \institution{Center for Democracy \& Technology}
  \city{Washington}
  \state{District of Columbia}
  \country{USA}
}
\email{awinecoff@cdt.org}
\authornote{Co-third authors contributed equally to this work.}
\author{Michael Madaio}
\affiliation{%
  \institution{Google Research}
  \city{New York}
  \state{New York}
  \country{USA}
}
\email{madaiom@google.com}

\authornotemark[1]
\author{Hoda Heidari}
\authornote{Co-last authors contributed equally to this work.}
\affiliation{%
  \institution{Carnegie Mellon University}
  \city{Pittsburgh}
  \state{Pennsylvania}
  \country{USA}
}
\email{hheidari@andrew.cmu.edu}
\author{Hong Shen}
\affiliation{%
  \institution{Carnegie Mellon University}
  \city{Pittsburgh}
  \state{Pennsylvania}
  \country{USA}
}
\email{hongs@andrew.cmu.edu}
\authornotemark[2]

\begin{abstract}

Model documentation plays a crucial role in promoting responsible AI (RAI) development. The emergence of Generative AI (GenAI) models has reshaped the conditions under which documentation is produced, particularly on open-source platforms where models are hosted and shared. To examine how these changes have manifested in developers' documentation practices, we interviewed 17 GenAI developers who document models on open-source platforms. Our findings illustrate that \textit{uncertainties} have become a defining feature of developers' GenAI documentation practices and that these uncertainties unfold in three interrelated forms: (1) normative and epistemic uncertainties in determining documentation content; (2) methodological uncertainties in evaluating and communicating model properties; and (3) ecosystemic uncertainties about who should document. We argue that these uncertainties in GenAI documentation require coordinated interventions, including infrastructural support to address epistemic and methodological uncertainties, community-based mechanisms to cultivate RAI documentation norms, and collaboration across supply-chain actors to address ecosystemic uncertainties.

\end{abstract}

\ccsdesc[500]{Human-centered computing~Empirical studies in HCI}
\keywords{Responsible AI, model documentation, open source}

\maketitle
\begin{table*}[!hb]
\centering
\begin{tabular}{p{0.2\linewidth} p{0.25\linewidth} p{0.45\linewidth}}
\toprule
\textbf{Dimension} & \textbf{Prior AI/ML Documentation Context} & \textbf{GenAI Documentation Context} \\
\midrule
\textbf{Nature of the model} &
Models built for narrow, bounded tasks with predictable and testable outputs \cite{Mitchell2019-cv, Arnold2018-wx}. &
Models have multi-task capabilities \cite{radford2019language} and emergent behaviors that cannot be evaluated with stable ground truth \cite{Wallach2024-id}, requiring evaluation of sociotechnical constructs that remain contested \cite{Weidinger2023-vz, Wallach2024-id, Aroyo2023-pk}. \\
\midrule
\textbf{Nature of the development context} &
Documentation created within relatively stable organizational workflows, with limited downstream modification \cite{Hind2019-jn, Heger2022-em}. &
Documentation produced on open-source platforms where open-weight models are continually finetuned, distilled, or quantized by diverse actors \cite{Kolides2023-rg, laufer2025anatomy}, often resulting in heterogeneous and unpredictable variants \cite{Ganguli2022-hp, Magooda2023-as, Qi2024-vo, Intelligence2024-io}. \\
\midrule
\textbf{Nature of governance} &
Model ownership situated within a single organization or team \cite{Hind2019-jn, Heger2022-em}. &
Responsibility diffuses across foundation model developers, finetuners, deployers, and platform intermediaries \cite{Yue2025-tv, Balayn2025-ca, Lee2023-bb, laufer2025anatomy, Bommasani2024-hy}; power dynamics have also shifted, with foundation model providers holding greater power and downstream actors having fewer resources \cite{Gray-Widder2023-xf, Burkhardt2024-rb}. \\
\bottomrule
\end{tabular}
\caption{
Drawing on previous literature, we identify three shifted conditions introduced by GenAI compared with prior AI/ML documentation context, which suggest a need to re-examine documentation practices.}
\label{tab:genai-motivation}
\end{table*}
\section{Introduction}

The responsible AI (RAI) community has long positioned AI documentation, such as Model Cards \cite{Mitchell2019-cv}, Datasheets \cite{Gebru2018-jt}, and FactSheets \cite{Arnold2018-wx}, as a foundational building block of robust AI governance, supporting transparency, accountability, and governance in AI systems \cite{Winecoff2025-yd,Liao2023-pc, Liao2023-fr}. These approaches have seen adoption in both open- and closed-source settings, with platforms like HuggingFace encouraging developers to provide structured model documentation \cite{hugging} and major AI companies such as Meta, OpenAI, and NVIDIA publishing documentation for their generative AI (GenAI) models \cite{OpenAI, Meta, NVIDIA}. As GenAI models proliferate, effective documentation has become even more crucial because these systems demonstrate novel and powerful capabilities, broad applicability, and enormous potential for downstream impact.

At the same time, the increasing prevalence of GenAI reflects a shift that challenges the underlying assumptions of existing documentation frameworks across three dimensions: the nature of the model, the nature of the development context, and the nature of governance (see Table~\ref{tab:genai-motivation}). First, unlike traditional ML models built for narrow, bounded tasks, GenAI models have open-ended and multi-task capabilities \cite{bommasani2021opportunities, Kapoor2024-al, radford2019language}, exhibit broader and novel risks \cite{Weidinger2023-od, Kapoor2024-al, Nist2024-fs}, and require evaluation of sociotechnical constructs that lack stable ground truth \cite{Xiao2024-mr, Ganguli2022-ox, Wallach2024-id, Grill2024-wx, hutchinson2022evaluation, Weidinger2025-gs}. Second, the contexts in which models are developed have shifted: open-source platforms now serve as critical infrastructure where GenAI developers host, share, and continuously modify models through fine-tuning, distillation, and quantization \cite{Srikumar2024-mh, bommasani2021opportunities, laufer2025anatomy}. Third, governance has become more fragmented: rather than residing within a single organization, responsibility now diffuses across supply-chain actors, including foundation model developers, finetuners, deployers, and platform intermediaries, with significant power imbalances among actors \cite{Gorwa2023-my, Widder2024-wk, Burkhardt2024-rb}.

Despite the importance of documentation in managing risks across the GenAI supply chain, prior research on documentation has not fully accounted for these sociotechnical shifts. Prior empirical work has investigated traditional ML practitioners' documentation practices in organizational settings, finding that practitioners generally understand what should be documented, but face barriers in execution due to ambiguous guidelines, workflow constraints, and insufficient organizational support \cite{Heger2022-em, Chang2022-fs, Hind2019-jn}. A separate body of HCI scholarship has examined documentation practices in open-source software communities, showing how developers rely on flexible, user-driven approaches and informal peer-to-peer knowledge sharing \cite{Geiger2018-bg, Dabbish2012-ao}. Yet the conditions introduced by GenAI open up new empirical terrain that neither body of work has sufficiently addressed: shifts in model properties change what can be known and measured about the model; shifts in development context may reshape the norms and expectations that guide documentation work; and shifts in governance structure may reconfigure how documentation responsibility is distributed and negotiated. As a result, we lack empirical understanding of how developers navigate documentation work under these evolving conditions. In order to fill this gap, we ask the following research question:

\textbf{Given the evolving conditions under which documentation is produced, 
how do GenAI developers document their models on open-source platforms, and what are the challenges they encounter?}


To answer this question, we conducted semi-structured interviews with 17 developers from various professional settings and roles with experience documenting GenAI models on open-source platforms.

Our findings reveal that uncertainty has become a defining characteristic of GenAI documentation work, manifesting in three distinct forms. First, developers face normative and epistemic uncertainties in determining documentation content, as they question whether responsible AI norms apply in open-source contexts and struggle to identify which model characteristics are feasible to measure and report. Second, developers encounter methodological uncertainties in evaluating and communicating model properties, as available model benchmarks and impact assessment methods often fail to capture GenAI's capabilities and risk profile. Third, developers navigate ecosystemic uncertainties in allocating documentation responsibility, as the fragmented GenAI supply chain distributes accountability across multiple actors with no clear norms for who should document bias, context-specific risks, or (un)intended model use cases.

These findings extend prior documentation scholarship in three key ways. First, we offer a conceptual lens---uncertainty---for understanding why GenAI documentation diverges from prior documentation work, articulating how shifts in model properties, development contexts, and governance structures generate distinct forms of uncertainty that existing documentation frameworks were not designed to address. Second, whereas prior empirical work characterized documentation challenges as primarily organizational \cite{Chang2022-fs, Heger2022-em}---practitioners know what to document but face workflow and incentive barriers, or find information gathering tedious---which presupposes that relevant information is known or knowable---we show that GenAI developers face more fundamental normative and epistemic uncertainties about what can and should be documented. Third, we offer design recommendations that move beyond simply adapting Model Cards for GenAI, proposing infrastructural support to navigate epistemic and methodological uncertainties, community mechanisms that recognize and incentivize documentation work through interpersonal learning, and collaborative frameworks with clarified documenter roles across supply chain actors.

\section{Related Work}

\subsection{The Shift from Traditional AI/ML to GenAI and Its Implications for Documentation}

Researchers have argued that the transition from traditional AI and ML systems to GenAI represents a fundamental shift as a new form of sociotechnical arrangement \cite{bommasani2021opportunities}, including changes in the nature of the model properties, the nature of the development context, and the nature of the governance structure (See Table \ref{tab:genai-motivation}). 

First, the rise of GenAI has fundamentally changed the nature of model properties. Whereas traditional supervised ML models typically operated within well-defined tasks and bounded output spaces, GenAI systems have far broader output spaces, can generate multimodal content \cite{Weidinger2023-vz}, and support a wide range of potential use cases \cite{Obradovich2024-vq, bommasani2021opportunities, Park2025-mz}. This expansiveness broadens the range of possible risks and harms, including the production of hallucinations and misinformation \cite{Li2025-hc, Weidinger2023-vz}. These risks are further amplified by the increased accessibility of GenAI technologies to both developers and end users \cite{Weidinger2023-vz, Solaiman2023-sa}.
A relevant shift concerns the evaluation approach to GenAI models. Traditional AI/ML models typically support evaluation on fixed tasks with bounded outputs, which allows
reporting performance metrics on a well-scoped task \cite{Mitchell2019-cv,Arnold2018-wx}. In contrast, recent work highlights that existing approaches for evaluating traditional ML systems are ill-suited for GenAI \cite{Jiao2024-eg, Wallach2024-id,
Ganguli2022-hp, Solaiman2023-sa, Magooda2023-as}. Unlike supervised models, GenAI systems must be evaluated on constructs (i.e., reasoning capability, toxicity, biases) that lack clear “ground truth” labels \cite{Wallach2024-id}. Further, these constructs are inherently sociotechnical; their definitions and interpretations are intertwined with social context, cultural norms, and domain-specific values \cite{Weidinger2023-vz, Wallach2024-id, Aroyo2023-pk}. 
As a result, GenAI evaluation reflects a distinct shift toward assessing open-ended, socially mediated constructs rather than measuring performance on predefined tasks, thereby reshaping what it means to characterize model capabilities and behaviors.

Second, this shift is also reflected in changes in the nature of development contexts for GenAI models in open-source environments \cite{bommasani2021opportunities, laufer2025anatomy}. Open-weight foundation models are being finetuned, distilled, quantized, and modified for diverse downstream purposes on open-source platforms \cite{Kolides2023-rg, laufer2025anatomy, Gorwa2023-my}. The scale of GenAI development in open-source is growing rapidly: recent analysis of 1.86 million models on Hugging Face revealed that text generation constitutes over 40\% of all model tasks, with some base models spawning sprawling ``family trees'' containing tens of thousands of derivative models across multiple generations of finetuning and adaptation \cite{laufer2025anatomy}. Meanwhile, such modifications can significantly alter a model's behavior, sometimes in unpredictable ways \cite{Ganguli2022-hp, Magooda2023-as, Qi2024-vo}. For example, one industry study found that certain finetuned variants were 21 times more likely than their parent models to produce toxic content \cite{Intelligence2024-io}. 

Third, GenAI systems have shifted the nature of model governance by complicating the AI supply chain and the power relationships between actors. New roles are being introduced, including foundation model developers, downstream finetuners, model distillers, and model deployers \cite{Yue2025-tv, Balayn2025-ca, Lee2023-bb}. In contrast to earlier ML workflows that typically assumed model development occurred within a single organization \cite{Hind2019-jn, Heger2022-em}, actors in these various roles now interact across organizational boundaries \cite{laufer2025anatomy, Bommasani2024-hy}, as downstream finetuners create layered dependencies by building upon and modifying models that may themselves be subsequently finetuned by other actors, resulting in complex, multi-tiered supply chains. Crucially, from a political economy perspective, scholars argue that this structure represents a fundamental departure from the power dynamics of traditional, task-specific ML models \cite{Gray-Widder2023-xf, Burkhardt2024-rb}. For instance, \citet{Burkhardt2024-rb} demonstrated that by functioning as “platform models” that necessitate an extreme concentration of data and computational resources, GenAI systems create distinct economic “moats” that centralize control; this shifts the supply chain relationship from one of independent development to one of infrastructural dependency, where downstream actors are increasingly reliant on a small oligopoly of foundation model providers.

These shifts raise questions about whether prior studies on documentation practice, conducted before the rise of GenAI, remain adequate for understanding documentation under these shifted conditions. Below, we review prior work on both ML documentation and open-source documentation, examining the assumptions underlying proposed frameworks and the empirical challenges reported in prior studies, and illustrate the research gap exposed by the shifted conditions introduced by GenAI.

\subsection{AI/ML Documentation as Responsible AI Artifacts}

The RAI community has long considered AI and ML documentation to be a key mechanism to improve transparency and accountability \cite{Liao2023-pc, Liao2023-fr, Mitchell2019-cv}. Publications focused on responsible AI at CHI, FAccT, and other venues have proposed documentation approaches including data documentation \cite{Gebru2018-jt, Jain2024-df, Bender2018-vy}, model documentation \cite{Mitchell2019-cv, Crisan2022-il, Mohammad2021-ns, McMillan-Major2021-ux, Cattell2024-gn, Kolt2024-zw}, and AI service documentation \cite{Arnold2018-wx}. These approaches aim to guide responsible usage, support ethical reflection during development, and strengthen governance mechanisms \cite{Winecoff2025-yd}.

Despite their potential value, empirical investigations have highlighted challenges in implementing documentation frameworks in practice, both through lab-based studies \cite{Liao2023-fr, Boyd2021-bs, Nunes2022-th, Nunes2024-zf, McMillan-Major2024-ma} and real-world investigations \cite{Yang2024-kd, Liang2024-hw, Heger2022-em, Chang2022-fs, Bhat2023-sv, Hind2019-jn}. Among these works, empirical investigations have primarily focused on traditional ML systems in organizational settings \cite{Heger2022-em, Chang2022-fs, Hind2019-jn}. For example, \citet{Heger2022-em} found that practitioners often struggle with determining documentation scope, audience, and level of detail. Similarly, \citet{Chang2022-fs} showed that insufficient incentives and limited organizational support commonly hinder the production of high quality documentation.
These studies characterize documentation challenges as primarily organizational, that is, practitioners generally understand what high quality documentation should include according to existing guidelines, but face barriers in execution due to ambiguity in guidelines, workflow constraints, and insufficient management support. However, these findings emerged from contexts where documentation work occurred within single organizations developing task-specific ML models. The shifted conditions introduced by GenAI---changes in the nature of the model, the development context, and the governance structure---suggest that organizational barriers alone may not fully account for the challenges developers face when documenting GenAI models in open-source environments.

\subsection{Documentation Practices in Open Source Communities: From Software to AI Models}

As GenAI documentation work now frequently occurs in open-source contexts, it is important to consider prior HCI scholarship on documentation happening in open-source communities. This body of work shows how developers and users participate in collaborative development, negotiate values such as transparency and accountability, and rely on infrastructures like discussion forums and platforms such as GitHub to coordinate work and sustain projects \cite{Dabbish2012-ao, Frluckaj2024-hh, Jahn2024-no, Barcellini2014-ux, Geiger2018-bg, Germonprez2018-pm, Filippova2015-bj, Jamieson2022-gf}. Open Source Software (OSS) refers to code licensed to enable anyone to freely access, modify, and repurpose it with minimal restrictions \cite{St-Laurent2004-zx}. OSS encompasses not only technical aspects of flexibility and accessibility, but also social dynamics that are often rooted in a hobbyist culture where distributed volunteers collaborate voluntarily \cite{Rossi2004-yk}. This culture is driven by personal enjoyment, a sense of freedom, and loosely connected networks of contributors \cite{Coleman2012-wg, Roberts2006-xr, Ahmad2021-rd}.

In OSS communities, documentation practices such as code documentation have evolved organically to serve the needs of developers \cite{Geiger2018-bg, Pawlik2015-qp, Arya2024-jb}. Prior literature \cite{Pawlik2015-qp, Pinho2024-jw, Kazman2016-im} highlights the user-driven and crowd-sourced nature of these practices, which commonly include project overviews, implementation details, usage guidelines, and terms of use. For instance, README files on Git-based platforms typically serve as the primary entry point for projects and contain these elements \cite{Aggarwal2014-em, Longo2015-fa}.

In recent years, open-source platforms have become central sites for GenAI model development. They now serve as critical infrastructure for model hosting and dissemination within the emerging GenAI supply chain \cite{Srikumar2024-mh, Ait2023-gt, Gorwa2023-my}. Open-source environments can increase access to AI development and potentially resist corporate concentration of power by leveling opportunities for individuals and smaller organizations \cite{Seger2023-ht, Kapoor2024-al, Howard2023-on}. However, open AI development can also align with corporate strategic interests \cite{Gray-Widder2023-xf, Srnicek2022-fp}, which can lead to open washing practices where claims of openness obscure gaps in meaningful transparency \cite{Gray-Widder2023-xf, Widder2024-wk, Liesenfeld2024-gj}. 

Given the central role of open-source platforms in GenAI development \cite{Osborne2024-tf, Widder2024-wk}, \citet{Gorwa2023-my} argue that these platforms have effectively become model marketplaces that require platform-level governance. Some platforms have responded by developing model documentation guidelines that aim to align with responsible AI principles \cite{hugging, Face2024-op}. Despite these efforts, recent analyses have revealed significant gaps in the transparency and completeness of open-source AI documentation artifacts \cite{Yang2024-kd, Liang2024-hw, Castano2023-tr}. For example, many documentation artifacts lack information about model limitations, intended uses, and potential social impacts \cite{Yang2024-kd}. These gaps often widen when base models are finetuned into new variants, which are frequently released with sparser documentation \cite{laufer2025anatomy}. As the need for detailed insight into development practices and risk profiles grows, the availability of such information often declines.

\subsection{Research Gap and Our Contributions}
A critical gap emerges across the prior literature: the shift from traditional ML to GenAI fundamentally changes the conditions under which documentation is produced. Prior scholarship on traditional AI and ML documentation, as well as work on open-source code documentation, offers valuable foundations for understanding documentation norms \cite{Geiger2018-bg}, organizational barriers \cite{Chang2022-fs, Heger2022-em}, workflow integration challenge \cite{Chang2022-fs, Hind2019-jn}, and insufficient management support \cite{Chang2022-fs}. This body of research characterizes documentation challenges as primarily \textit{organizational barriers} in ML contexts---practitioners understand what to document but face execution barriers \cite{Chang2022-fs, Heger2022-em}---or as \textit{motivational} in open-source contexts, where documentation is often deprioritized as less enjoyable than coding work \cite{Geiger2018-bg}. Both framings presuppose that relevant model information is knowable and that established practices exist from which developers can report results. However, this work largely examines documentation within single organizations \cite{Heger2022-em, Hind2018-jc} or open-source communities focused on software libraries \cite{Geiger2018-bg}, and therefore does not account for the shifted conditions GenAI introduces.

By examining GenAI model documentation in open-source settings, this study contributes to prior documentation scholarship by revealing challenges that are fundamentally different in kind, not merely degree. Our findings illustrate that uncertainty has become a defining feature of GenAI documentation. By articulating these uncertainties, our study offers a conceptual lens for understanding why GenAI documentation departs from prior practices, clarifies conditions that future frameworks must address, and offers design recommendations for platforms and the RAI community.

\section{Methods}

In this study, we examine how GenAI developers approach model documentation on open-source platforms. Following \citet{Gorwa2023-my}, we define ``open-source platforms'' as both general-purpose software development platforms that host AI models (e.g., GitHub) and specialized AI model marketplaces (e.g., Hugging Face) where users can upload and share AI models and AI-related datasets that others can download, modify, and build upon. We define ``GenAI developers'' broadly to include anyone involved in creating, adapting, or deploying GenAI models, and, where appropriate, we mention their specific roles across the development pipeline \cite{Kolides2023-rg,Jiang2023-ev}. Following \citet{Mitchell2019-cv}, we define ``model documentation'' as publicly available materials that report basic information, training data, evaluation results, and intended use cases for a model. Model documentation can encompass a variety of artifacts, such as Hugging Face Model Cards, GitHub README.md files, or technical reports.

\begin{table*}[!h]
\centering
\small
\setlength{\tabcolsep}{2pt} 
\renewcommand{\arraystretch}{0.95}
\begin{tabular}{@{}p{0.04\linewidth}@{\hspace{1pt}}p{0.25\linewidth}@{\hspace{1pt}}p{0.20\linewidth}@{\hspace{1pt}}p{0.20\linewidth}@{\hspace{1pt}}p{0.12\linewidth}@{\hspace{1pt}}@{}p{0.15\linewidth}@{}}
\toprule
\textbf{ID} & \textbf{Roles} & \textbf{Professional Context} & \textbf{Development Focus} & \textbf{Platforms} & \textbf{Model Reuse Records}\\ 
\midrule
P1 & Model distillation & Industry  & Efficient LLM & HF & > 1k per month\\

P2 & Model fine-tuning \& Deployment  & Industry \& Academic & Privacy-preserving LLM & GitHub, HF & 100 - 1k per month \\

P3 & Model fine-tuning & Academic \& Individual & LLM for research applications & HF & < 100 per month \\

P4 & Foundation model & Academic & Pre-trained LLM & HF, GitHub & > 1k per month\\

P5 & Model fine-tuning \& Distillation & Industry \& Individual & Efficient LLM & HF & > 1k per month\\

P6 & Model fine-tuning \& Deployment & Industry \& Individual & LLM for creative writing & HF, GitHub & > 1k per month \\

P7 & Model fine-tuning & Industry & Visual language model & GitHub, HF & > 1k per month\\

P8 & Model fine-tuning \& Deployment & Academic & LLM for Survey analysis & HF & --  \\

P9 & Model fine-tuning & Individual \& Industry & Text-to-image model & GitHub, HF & -- \\

P10 & Model fine-tuning & Individual & LLM for code generation & HF & 100 - 1k per month \\

P11 & Model fine-tuning & Individual \& Industry & -- & GitHub & --\\

P12 & Model fine-tuning & Individual & Text-to-speech for gaming & HF, GitHub & 100 - 1k per month \\  

P13 & Foundation model & Industry & -- & HF & > 1k per month \\ 

P14 & Model fine-tuning & Academic & LLM for Mental Health & HF & < 100 per month \\

P15 & Model fine-tuning & Individual & LLM Agents & HF & 100 - 1k per month \\

P16 & Model fine-tuning & Academic & LLM for Text Classification & GitHub & -- \\

P17 & Model fine-tuning & Industry \& Academic & LLM for Enhanced Reasoning & GitHub & -- \\
\bottomrule
\end{tabular}
\caption{Participant roles and model development activities. Our participants engaged in various roles in GenAI model development, including fine-tuning existing models, distilling larger models, pre-training LLMs, and building applications that integrate these models \protect\citep{Kolides2023-rg,Jiang2023-ev}. Note: HF = Hugging Face. Model reuse records were collected from Hugging Face model download profile page where available. }
\label{tab:participants}
\Description{This table summarizes participant roles and model development activities for 17 participants (P1-P17). The table has five columns: ID, Roles, Professional Context, Development Focus, Platforms, and Model reuse records. Participants' roles include model distillation, fine-tuning, deployment, and foundation model development. Professional contexts span industry, academic, and individual settings, with many participants working across multiple contexts. Development focuses include efficient LLMs, privacy-preserving LLMs, research applications, creative writing, visual language models, survey analysis, text-to-image models, code generation, text-to-speech for gaming, mental health applications, LLM agents, text classification, and enhanced reasoning. The primary platforms used are Hugging Face (HF) and GitHub, with most participants using one or both platforms. The participants represent diverse backgrounds in GenAI model development, encompassing various technical approaches from fine-tuning existing models to pre-training foundation models and building integrated applications.}
\end{table*}

\subsection{Study Protocol}

Following established qualitative research methods for semi-structured interviews \cite{patton2014qualitative}, 
we designed our protocol to begin with broad, open-ended questions about participants' background in GenAI 
development and their past experiences with model documentation, examining their motivations, existing practices, challenges, tools and resources they used for documentation, as well as their approach to navigating different documentation sections and team collaboration. We maintained flexibility to probe deeper into specific 
areas through follow-up questions, particularly around dimensions that warranted deeper investigation based on 
synthesis of prior literature on the challenges brought by shifted conditions from traditional ML models to GenAI models, which we summarized in Section 2.1.

To investigate aspects unique to GenAI, we specifically designed questions that explored how developers document emergent model behaviors, how they decide what to disclose when evaluation lacks stable ground truth, and how they navigate responsible AI expectations in open-source contexts. For example, when 
participants described the reporting process for evaluation results, we probed how they decided which parts of their evaluation 
design and results to document, how they selected benchmarks, and how they approached documenting model 
capabilities versus risks. When they mentioned collaboration, we asked follow-up questions about how they think of responsibility 
distribution across team members or supply chain actors. This adaptive questioning approach allowed us to 
investigate areas highlighted by prior literature while remaining open to challenges and navigation strategies that emerged organically from developers' experiences.

 Each interview lasted between 30 and 65 minutes. All interviews were recorded and transcribed for analysis with participants' consent. Each participant was compensated \$60 USD.  The study received approval from the Institutional Review Board (IRB) at the first author's institution.

\subsection{Data Collection}
To capture a range of perspectives from the open-source GenAI development ecosystem, we adopted a purposive sampling approach \cite{patton2014qualitative}, recruiting via multiple channels. 
This included leveraging personal and professional networks (e.g., Slack channels), posting messages on social media (e.g., LinkedIn, Discord\footnote{We posted on Discord channels that open-source communities use for discussion and community-building.}), and direct outreach to developers on open-source platforms, such as Hugging Face and GitHub. For recruitment via Hugging Face, we specifically reached out for developers of popular GenAI models (sorted by download counts) to ensure participants had experience documenting models intended for downstream reuse. Participants were recruited based on the following criteria: (1) experiences in documenting GenAI models they uploaded on open-source platforms; 
(2) age 18 or older; (3) located in the US or EU, due to IRB restrictions.


We interviewed 17 participants involved in various stages of GenAI development---including foundation model development, fine-tuning, distillation, and deployment \cite{Kolides2023-rg,Jiang2023-ev}. Our participants worked in a range of professional settings, including academia, industry, and independent open-source communities. This diversity of experiences helped us capture multiple perspectives across the GenAI supply chain. 
We list demographics in Table~\ref{tab:participants}. In addition, we report monthly download rates for the models hosted on Hugging Face (as of Dec 2025) as a proxy for model reuse, if applicable.

To complement our interviews, we also collected publicly available documentation materials produced by our participants. We gained consent from 9 out of 17 participants to analyze the documentation artifacts. Relevant documentation content included model descriptions, training data information, evaluation metrics, intended use cases, and sections addressing bias, risks, and limitations.

Interview data were anonymized using randomly-assigned alphanumeric 
identifiers (e.g., P1, P2). For documentation artifacts, we replaced all identifying information including company 
names, model names, and other identifiable details with generic placeholders (e.g., [COMPANY NAME], Model 1, 
Model 2). Additionally, we rephrased any sentences that we identified as potentially searchable or traceable to 
their original sources, while preserving the semantic content and structure of the documentation. This approach 
allowed us to present authentic examples of documentation patterns while protecting participants' identities and 
their proprietary work.

\subsection{Data Analysis}
We employed a reflexive thematic analysis approach \cite{braun2006using} to analyze the interview data.
The first author performed an initial open coding of all the interview transcripts and kept detailed memos throughout the coding process to document emerging patterns and potential themes. The codebook was developed iteratively by the first author through this open
coding process, beginning with descriptive codes grounded in the data and progressively developing more analytical
codes. 
The research group also held regular collaborative sessions to discuss emerging codes, resolve discrepancies
and ambiguities in the codes, and iteratively refine our coding scheme. We then generated higher-level themes that captured patterns of shared meaning across all the codes via both inductive and deductive coding, while keeping our research questions in mind. The research team reached shared interpretive agreement through these discussions, consistent with the tradition of reflexive thematic analysis \cite{braun2006using, mcdonald2019reliability}. Through this iterative process, we identified several key themes related to GenAI developers' documentation practices and challenges in open source environments. 
Ultimately, this process generated 131 first-level themes, 9 second-level themes, and 3 third-level themes.

After identifying themes from the interview analysis, the first author applied them to examine documentation artifacts. 
Documentation materials were segmented into excerpts based on sections outlined in documentation guidelines \cite{Mitchell2019-cv} (e.g., model capabilities, risks and limitations, intended use cases, evaluation results).  The first author then identified common documentation patterns across these excerpts, triangulating how the uncertainties identified in the interview data materialized in concrete documentation examples. 
This analysis revealed five distinct documentation patterns that we present as exemplar excerpts in Section 4 (Figures~\ref{fig:model-1-card}--\ref{fig:comprehensive-risks}).

We present our results and first-level themes in Section 4. All second and third-level themes are presented in the Appendix. Using the standard practice of reflexive thematic analysis, we did not calculate the reliability between coders, as repeated discussions about
discrepancies are integral to generating the codes and themes \cite{braun2019reflecting, mcdonald2019reliability}. 



\section{Findings}
\begin{table*}[!t]
\centering
\caption{Our findings identify three forms of uncertainty introduced by the shift from traditional ML to GenAI, each reshaping documentation practices in distinct ways.}
\label{tab:uncertainty}
\begin{tabular}{p{0.2\linewidth} p{0.35\linewidth} p{0.35\linewidth}}
\toprule
\textbf{Form of Uncertainty} & \textbf{Definition (Conceptual Lens)} & \textbf{How It Appears in Developer Documentation Practice} \\
\midrule
\textbf{Normative and epistemic uncertainty} & 
Uncertainty in determining documentation content: \textit{normative uncertainty} concerns whether RAI norms apply in open-source contexts; \textit{epistemic uncertainty} concerns what model characteristics are knowable or measurable & 
Developers question whether RAI norms apply; privilege technically tractable characteristics over important but hard-to-measure properties; hesitate on disclosure scope amid conflicting priorities \\
\midrule
\textbf{Methodological uncertainty} & 
Uncertainty in \textit{how} to evaluate and communicate model properties, given under-defined procedures, limited benchmarks, and lack of established reporting conventions & 
Developers lack clear procedures for evaluation design, result interpretation, or translating assessments into meaningful documentation; rely on proxy metrics despite doubts about validity \\
\midrule
\textbf{Ecosystemic uncertainty} & 
Uncertainty in \textit{who} should document, as responsibility fragments across a supply chain with partial visibility and no clear accountability structures& 
Documentation becomes negotiated and distributed; developers shift accountability upstream (to foundation model providers) or downstream (to deployers); gaps persist across supply chain boundaries\\
\bottomrule
\end{tabular}
\end{table*}
Through our analysis, we find that uncertainty, rather than implementation barriers, has become a defining feature of how GenAI developers document models on open-source platforms. We identify how GenAI developers navigate three forms of uncertainty that depart from prior accounts of documentation work in traditional ML settings. First, developers face normative and epistemic uncertainties in determining documentation content. Second, they grapple with methodological uncertainties in how to evaluate and communicate model properties. Third, developers encounter ecosystemic uncertainties around how documentation responsibilities should be allocated. 
\begin{figure*}[!hbp]
\centering
\small
\begin{minipage}{0.8\textwidth}
\begin{graybox}{Model 1 Model Card}

\textbf{Model Description}

This is an enhanced version of the base model developed by Research Team [RESEARCH TEAM NAME]. The training dataset incorporates multiple specialized sources focused on domain-specific applications. For inquiries, please contact [COMPANY NAME].

\textcolor{red}{\textbf{Warning:}}  This system is not appropriate for users under 18. Output may include mature themes and adult-oriented material.

\vspace{0.1cm}

\textbf{Training Data}

The training corpus consists of [X] distinct datasets. [further details redacted]

\vspace{0.1cm}

\textbf{Usage Instructions}

This model can be used directly with standard text generation pipelines. Example implementation:

\begin{lstlisting}[language=Python, basicstyle=\footnotesize\ttfamily, frame=single]
from transformers import pipeline
generator = pipeline('text-generation', 
                    model='model-1-[X]b-version')
output = generator("Your prompt here", 
                  max_length=[XXX])
\end{lstlisting}

\vspace{0.1cm}

\textbf{Limitations and Biases}

As with all large language models, this system may exhibit biases related to demographic factors, professional domains, and cultural contexts. 

\vspace{0.1cm}

\textbf{License}

Model 1 is released under the Research Community License v[X].[X], Copyright (c) [COMPANY NAME]. All Rights Reserved.

\end{graybox}
\end{minipage}
\caption{Example of minimal model documentation that adheres to basic platform requirements while relying on alternative communication channels.}

\label{fig:model-1-card}
\Description{This figure shows a model card for "Model 1" presented in a gray-bordered text box. The model card contains six main sections: Model Description explains this is an enhanced version of a base model by a research team, with a red warning that the system is not appropriate for users under 18 due to mature content; Training Data mentions the corpus consists of multiple distinct datasets with further details redacted; Usage Instructions provides Python code showing how to use the model with the transformers library pipeline for text generation; Limitations and Biases acknowledges potential biases related to demographic, professional, and cultural factors; and License states the model is released under a Research Community License. The model card demonstrates minimal documentation that meets basic platform requirements while suggesting additional details are available through other communication channels.}
\end{figure*}

\subsection{Normative and Epistemic Uncertainties in Determining the Documentation Content}

\label{contentuncertainties}


\subsubsection{Uncertainty in Moving Beyond Established Open-Source Documentation Practices}

While prior work has shown that developers’ documentation practices are often influenced by organizational norms and culture, our findings show that developers documentation practice is influenced by a \textit{normative uncertainty} specific to open-source contexts. Rather than treating RAI guidelines as directly applicable, developers questioned whether these norms align with the open-source community’s longstanding preference for informal, lightweight, and highly flexible documentation. This uncertainty shaped their hesitation to expand or formalize documentation content.


Developers’ uncertainty about expanding documentation often reflected their own struggles with using it, as they tended to favor the informal, flexible style of open-source knowledge sharing instead of the structured demands of RAI documentation. They often found that using lengthy, standardized documentation artifacts conflicted with their preferred nature of agile open-source development.
{Some developers expressed their preference for informal, flexible open-source knowledge sharing over lengthy, standardized RAI documentation that conflicted with agile open-source development practices.} P9, who is an active contributor on Hugging Face, notes: \textit{``They do have good document[ation], but specifically for the model cards, ... you literally have to read a full, a very long page... The time and the learning curve to just spend on any syntax or semantics is just a little too much.''} Perusing a lengthy documentation artifact was a less desirable method for learning about model functionality than reading abbreviated documentation and experimenting with the model:  \textit{``I enjoy flexibility as a programmer ... [For model documentation,] it's an initial learning curve, but still it's a learning curve.''}

Similarly, developers felt many users would prefer to rely on informal communication methods available within open-source communities and questioned whether users of their models actually valued or wanted comprehensive documentation. For example, P6 observed that users of his model often bypass formal documentation entirely, instead seeking quick recommendations through channels like Discord. Using such channels in lieu of model documentation, open-source users often reach out to the model creator and ask questions like: \textit{``what does the model do? Is there anything else that I need to know about the model?''} This pattern reflects the traditional hobbyist preference for direct, peer-to-peer knowledge sharing in open source communities. As P6 explained, developers in this community prefer \textit{``word of mouth, mouth-to-mouth talk.''}

Limited community engagement led many developers to question the value of RAI documentation efforts. Specifically, some questioned whether it is meaningful to invest time in communicating information that is not traditionally required by open-source developers and users. This perception was reinforced by minimal community engagement from RAI-focused sections of their documentation. P4, for example, noted that despite extensively documenting ethical considerations, he never received feedback on the bias and limitations section from the open-source community, making him feel that \textit{``no one would've engaged with the model [documentation] in its entirety, even though every bit of it is important.''} Such experiences strengthened developers' beliefs that these efforts offered limited value.  

Similarly, P7 reflected on his team's efforts in documenting ethical considerations, including removing opted-out content from their training dataset and conducting red-teaming exercises. However, he observed a lack of community engagement in his model card's sections on ethical considerations despite his team's efforts: \textit{``Unfortunately... we never really had any positive feedback on this [section on bias and limitations]. I think mostly people will say that it is good, but they don't find it necessary.''} 

As a result, these uncertainties are manifested in participants' documentation artifacts---some of our participants tended to create minimal documentation that adheres only to basic platform requirements while assuming others will use informal communication channels within developer communities. For example, Figure \ref{fig:model-1-card} shows an example model card that only includes standard sections like model descriptions and usage instructions, but deliberately omits comprehensive RAI components such as detailed bias assessments, thorough limitation discussions, or extensive evaluation results.

Still, individual and organizational-level advocacy was sometimes effective in promoting more comprehensive documentation, particularly documentation that addresses RAI concerns. P4 shared that his team strongly valued multicultural inclusion: \textit{``There was a very intentional decision of having people focus on low-resource languages.''} As a result, the team documented both their efforts and limitations in achieving this goal, such as by noting when certain language families remained underrepresented. Individual leadership was also a key factor---P4 described how having someone dedicated to climate impact led to systematic documentation of environmental effects within his team.



\subsubsection{Uncertainties around What Model Characteristics are Feasible to Measure and Report Faithfully} \label{whattomeasure}


When documenting sections such as model evaluation and model biases and risks, our participants also expressed uncertainty about which model characteristics are feasible to measure and report. We find that this uncertainty operates on two related epistemic layers. First, participants questioned whether a given behavior or risk is measurable at all, leading them to privilege characteristics that seem technically tractable or easy to quantify. Second, they questioned whether an available benchmark provides a legitimate or accepted way of measuring that characteristic, which pushed them toward reporting only results tied to widely adopted benchmarks. P7 cited technical feasibility as a deciding factor in choosing to evaluate his multi-modal model on FairFace \cite{karkkainenfairface}, an established benchmark for detecting gender, race and age bias: \textit{``it is pretty easy to run the evaluation. You just need to integrate that into the code.''} He also emphasized that FairFace's broad adoption enabled easy comparison: \textit{``it was used by other people, so that in this way you can compare your model against them.''} However, he did not choose to evaluate and report hallucinations in his models because he \textit{``rarely sees companies using such benchmarks.''} He therefore concluded that hallucinations are a \textit{``very hard problem to solve''} in his model domain.

Similarly, P3 also heavily prioritized technical feasibility and quantifiability in evaluation and documentation. He stated: \textit{``if one were to pick a factor [to evaluate], I would pick whichever is the easiest and most quantifiable to document.''} P3 felt it was worthwhile to document the model's reliability because model reliability is \textit{``studied really nicely and there already exist quantitative measures for it.''} In contrast, he viewed model toxicity or bias as \textit{``way too hard''} to document because \textit{``bias and toxicity are like sarcasm. Sarcasm across cultures is different. It is just multidimensional... there is cultural angle to it, there's geographic angle to it... race could also become a problem.''} Although research shows that toxicity and bias can be measured across cultures \cite[e.g.,][]{Davani2024-rr}, developers' documentation decisions for such evaluations appear driven more by \textit{perceived} feasibility than by the actual feasibility or importance of these risks. 

This approach reveals a concerning disconnect between what participants consider feasible to report and what constitutes a comprehensive documentation of model capabilities and limitations. Participants often prioritize ``documentability'' based on the availability of quantifiable measurements or established benchmarks, while dismissing crucial issues such as model bias and toxicity as too challenging to document. 
A key factor underlying this pattern is the lack of a single ground truth for many GenAI outputs: unlike traditional classification tasks with fixed labels, evaluating phenomena such as hallucination, bias, or toxicity often requires subjective judgments across contexts, which developers feel especially uncertain to grapple with. We return to this in the Discussion, as contested ground truth \cite[e.g.,][]{muller2021designing} and rigorous application of subjective judgments \cite[e.g.,][]{miceli2020between} are a hallmark of much of HCI research, and are not technically infeasible as such. 


\subsubsection{Uncertainties around Appropriate Levels of Disclosure Amid Conflicting Priorities}

Beyond epistemic uncertainties about what can be known or measured, participants also faced normative uncertainty about how much detail they should disclose when documentation involves competing ethical, legal, and commercial priorities. This led to hesitation about the appropriate scope and specificity of disclosures.

When documenting training data, some express uncertainty about the boundaries of what can be shared since their datasets might contain unlicensed data: \textit{``some of my training data is copyrighted... [which means that] for research purposes, you can use the dataset, but if you distribute it, you basically can get into legal issues.''} As a result, developers often choose to limit their documentation as a strategy for managing legal risks. For example, while they might include basic information, such as dataset size, they deliberately withhold more detailed information that might reveal any legal violations associated with their data usage.

For participants working in industry, they expressed uncertainty about how much they can disclose regarding proprietary datasets, as companies increasingly consider data as a valuable form of intellectual property.  P9 articulates how withholding certain details of their datasets can be a means of defending a competitive advantage: \textit{``if you release the dataset... it might be a little bit too open for our competitors to build a competing product.''} He further mentioned some cases where they had to take the dataset out of the open-source environment, once they realized the dataset could be monetized by others and used for commercial purposes: \textit{``the thing is even though we are making some of the applications open, it doesn't necessarily mean that the company would be forgoing all the monetary benefits out of it.''} Similarly, P17 described how her company requires VP approval to publish a dataset, noting a shift in industry practices: \textit{``compared with academics, industry now holds more and more secrecy towards the dataset and the model.''}

We also observed that some participants feel uncertain about how much they should provide when documenting considerations around the model's risks, biases, and limitations, as this conflicts with marketing interests. Marketing considerations push developers to emphasize their models' strengths and potential, while ethical considerations demand transparent and holistic disclosure of risks and limitations. For instance, when interviewing P3 about why he did not include detailed limitations in his documentation, he expressed his perception on the open-ended capabilities of GenAI models: \textit{``I think it's understood that large language models are non-deterministic in nature, right?''} He then asked \textit{``why would anyone want to share their weaknesses [when you can selectively praise the model's potential]?''} Similarly, some participants acknowledge that they could conduct safety evaluations, but chose not to prioritize them, instead favoring marketable metrics. P1, who is working on a startup that provides lightweight and efficient GenAI models, admitted that he is limiting the reporting on bias and limitations: \textit{ ``I think it is technically possible to measure hallucinations or bias... I believe there are some benchmarks...''}. However, he indicated that his company's focus on marketable efficiency metrics took precedence: \textit{``We are first to focus on efficiency and then we can do more things about whether these models could say something which is wrong...''} 

These uncertainties resulted in our participants' conservative disclosure practices, which reflects a broader challenge in the current GenAI ecosystem: while market dynamics drive rapid model development and deployment, documentation standards have not yet evolved to provide sufficient granularity to guide complex disclosure decisions. Although existing documentation guidelines specify what aspects to report, they often lack practical guidance on navigating competing interests and risks in different contexts. As a result, developers typically default to minimal disclosure, though this risk-averse strategy may ultimately hinder the adoption of responsible GenAI documentation practices.

\subsection{Methodological Uncertainties in Evaluating and Communicating Model Properties}

\label{Howtoreport}

While the previous section highlighted normative and epistemic uncertainties, participants also faced methodological uncertainties regarding how to document model properties in practice. Developers described evaluation for GenAI as procedurally under-defined: existing tools and benchmarks capture only narrow slices of model characteristics, established reporting conventions from traditional ML rarely apply to diverse, open-ended tasks, and few structured methods exist for assessing downstream uses or context-specific risks. As a result, even when developers had a sense of what should be documented, they lacked clear procedures for designing evaluations, interpreting results, or translating those assessments into documentation that would be meaningful for diverse downstream users.

\subsubsection{Uncertainties in How to Report Model Capabilities}
When documenting models' capabilities, participants expressed that they often feel uncertain about how to accurately report them due to 
the limitations of existing benchmarks. As a result, they often choose to rely on some ``proxy metrics,'' despite concerns about their actual usefulness.  



Participants who work on foundation models shared that the lack of industry-wide standard evaluation benchmarks creates uncertainty about how to present model capabilities in a transparent and comparable way. As P13, noted: \textit{``everyone tests their model on different tasks.''} He described how some competitors selectively report results for specific metrics (e.g., only using IFEval's loose criteria while omitting the strict criteria\footnote{IFEval is a benchmark to evaluate a language model's capability to follow instructions \cite{Zhou2023-gv}. It has two different metrics: strict accuracy checks if instructions are followed exactly, while loose accuracy is more flexible.}), potentially creating misleading comparisons: \textit{``it's difficult to present [results] in a fair way.''} Some also worried that existing benchmarks may become inadequate due to data or concept drift.\footnote{Data drift refers to changes in the distribution of input data over time, after a model has been deployed} Similarly, P1 mentioned that static benchmarks cannot adequately capture the model's performance on dynamic data when models are deployed in real-world settings: \textit{``even if there is one questioning-answering dataset, maybe later in 10 years there will be different test datasets... there will be a data shift, then the LLM (performance) will not be good anymore.''} 

Relatedly, those who fine-tune models for a specific purpose also feel uncertain about how to report the model's capability because benchmarks often do not exist for the specific use cases they care about. That is, even if they \textit{do} know what the specific use case is for their model, they may be unsure of how to document performance on that use case given limitations in the available evaluation methods. P6, who develops GenAI for storytelling, highlighted how existing standardized, generalized evaluation metrics such as question answering (QA) benchmarks, became meaningless in his model's use case: \textit{``We want to test our models on how well they can write certain content and how good that content is. And that's more difficult to do with existing dataset, as there is no good benchmark for storytelling.''} This participant feels they can only conduct the evaluation by observing output using the same prompt to see if the model has improved. As a result, he chose to omit standardized benchmark results from his model card, explaining: \textit{``why do I need it? You don't need math QA benchmarks for story writing.''} 

Faced with a dearth of appropriate specific benchmarks on how to report model capability, many chose instead to rely on ``proxy metrics'' -- widely used benchmarks that provide a general understanding of the model’s overall capabilities (e.g., ``reasoning''). Despite the fact that P1 didn't entirely trust the leaderboard benchmarks because they \textit{``[don't] give a full complete view on the LLM capacity,''} he still thought such metrics could prove useful in giving users \textit{``some first insights about the model's capability.''} These metrics could at least give users a general impression of, for example, \textit{``Is that a very bad LLM or good LLM?''}. However, such proxies may not always be robust measures of the construct that developers or users care about \cite[cf.][]{diaz2024scaling,jacobs2021measurement}.

\begin{figure}[!htbp]
\centering

\small
\begin{graybox}{Benchmark Results - Model 3}

\vspace{0.1cm}

\begin{center}
\small
\begin{tabular}{|l|c|c|}
\hline
\textbf{Model} & \textbf{TruthfulQA} & \textbf{ReasoningBench} \\
\hline
Model 3 & \textbf{XX.XX\%} & \textbf{XX.XX\%} \\
Baseline Model A  & XX.XX\% & XX.XX\% \\
Baseline Model B  & XX.XX\% & XX.XX\% \\
Baseline Model C & XX.XX\% & XX.XX\% \\
\hline
\end{tabular}

\vspace{0.1cm}

\begin{tabular}{|l|c|c|}
\hline
\textbf{Model} & \textbf{KnowledgeEval} & \textbf{SafetyBench} \\
\hline
Model 3 & \textbf{XX.XX\%} & \textbf{XX.XX\%} \\
Baseline Model A & XX.XX\% & XX.XX\% \\
Baseline Model B & XX.XX\% & XX.XX\% \\
Baseline Model C & XX.XX\% & XX.XX\% \\

\hline
\end{tabular}
\end{center}

\end{graybox}

\caption{Example of benchmark results presentation that demonstrates how developers use generic benchmark results to present their model's capabilities, while providing limited contextual information on how to interpret these benchmark results in specific use-cases. }
\label{fig:benchmark-results}
\Description{This figure displays benchmark results for "Model 3" in a gray-bordered box containing two performance comparison tables. The first table compares four models (Model 3 and three baseline models A, B, and C) across TruthfulQA and ReasoningBench metrics, with Model 3's scores highlighted in bold and showing placeholder "XX.XX\%" values. The second table uses the same four models to compare KnowledgeEval and SafetyBench performance, again with Model 3's results emphasized in bold. All numerical values are redacted as "XX.XX\%" placeholders. The figure illustrates how developers present benchmark comparisons to showcase their model's capabilities while providing minimal contextual information about interpreting these results for specific applications.}
\end{figure}

The challenge of documenting model capabilities becomes apparent when examining how participants present evaluation results in their documentation artifacts. Fig \ref{fig:benchmark-results} exemplifies a common pattern we observed: developers include quantitative performance metrics to demonstrate their model's capabilities, but provide limited contextual information about what these benchmarks actually measure or how the results should be interpreted by potential users. This artifact represents the broader tension participants face between wanting to demonstrate model competence through standardized metrics and their uncertainty about whether these metrics truly capture the capabilities that matter for their specific use cases.



\subsubsection{Uncertainties in How to Report Intended and Unintended Model Use Cases}\label{How to report intended and unintended use case}

Developers struggle to define and report the scope of potential uses for their GenAI models for two reasons. First, GenAI systems can produce a vast generative output space compared with traditional ML models, making it difficult to anticipate the full scope of potential downstream applications, and second, pre-trained foundation models can be applied to a variety of downstream use cases and contexts \cite{bommasani2021opportunities}. The challenge is further amplified in open-source environments, where developers have limited visibility into how their models are actually deployed downstream. Faced with these uncertainties, we found that developers typically respond by seeking guidance from domain specialists such as legal advisors to help establish boundaries around appropriate use, or they provide minimal documentation about intended and unintended use cases, expecting downstream users to be accountable for the model impacts. 

When discussing documentation of intended use cases, developers highlighted the primary challenge posed by the open-ended, general-purpose  nature of GenAI models, which makes it difficult to establish clear boundaries around appropriate use cases, as \textit{``there are too many different possibilities.''} P9 illustrated this uncertainty by comparing GenAI to his previous work on a traditional ML model for empathy detection from text: \textit{``In that instance, you can pretty much say... the use of the model can be determined properly.''} He contrasted this experience with the attempt to clearly define intended use with generative systems: \textit{``Think about the large language models, it's much more generalized. You can't specifically say it should be used for this purpose.''} 

Defining unintended use cases was equally challenging for developers, especially on open-source platforms. With limited tools to comprehensively assess potential misuse scenarios, identifying unintended uses often relies on subjective intuition. According to P9, an industry practitioner, he has limited visibility into how his models may be used inappropriately, as the platform is open and \textit{``basically everyone can download it... it is much more [of] a speculative guess of how [the model] could ever be misused.''} He noted that although he tried to get some insight into potential misuse through internal testing with his colleagues, this informal approach did not feel sufficiently systematic or comprehensive. As a result, he often turned to legal consultants in his company because they: \textit{``told me that sometimes people misuse it in certain ways, which may not be in line with what I'm thinking in the first place.''}  

\begin{figure}[!tbp]
\centering
\small
\begin{graybox}{Model 2 - Usage Restrictions}

\textbf{Misuse and Out-of-scope Use}

This model is unsuitable for \textbf{high-risk} environments and critical decision-making contexts. It lacks the reliability required for scenarios with significant consequences affecting people's lives, careers, or personal welfare. While model outputs may seem accurate, they cannot be guaranteed as factually correct.

\vspace{0.1cm}

\textbf{Out-of-scope uses include:}
\begin{itemize}
    \item Usage for assessing individuals, such as for employment, education, or credit
    \item Applying the model for crucial automatic judgments, generating factual content, creating summaries that require reliability , or generating predictions that must be correct
\end{itemize}

\vspace{0.1cm}

Deliberate deployment of the model to cause damage, violate \textbf{human rights}, or engage in other forms of harmful conduct constitutes improper use.

\vspace{0.1cm}

\textbf{This includes:}
\begin{itemize}
    \item Disparagement and defamation
    \item Generating spam
    \item Disinformation and manipulation
    \item Harassment 
    \item Deception
    \item Unconsented impersonation and imitation
    \item Unconsented surveillance
\end{itemize}

\end{graybox}

\caption{Example of comprehensive documentation of potential misuse scenarios and usage boundaries.}
\label{fig:usage-restrictions}
\Description{This figure shows usage restrictions for "Model 2" in a gray-bordered text box. The document is divided into two main sections: "Misuse and Out-of-scope Use" explains that the model is unsuitable for high-risk environments and critical decision-making contexts, emphasizing that outputs cannot be guaranteed as factually correct, followed by a bulleted list of prohibited uses including individual assessments for employment/education/credit and critical automatic judgments requiring reliability. The second section describes deliberate harmful deployment as improper use that violates human rights, followed by a bulleted list of seven specific prohibited activities: disparagement and defamation, generating spam, disinformation and manipulation, harassment, deception, unconsented impersonation and imitation, and unconsented surveillance. The figure demonstrates comprehensive documentation of potential misuse scenarios and usage boundaries.}
\end{figure}

While many participants struggled with defining appropriate use cases for their GenAI models, some did invest significant effort in documenting potential misuse scenarios and establishing clear usage boundaries. Figure \ref{fig:usage-restrictions} represents one of the more comprehensive approaches we encountered, demonstrating how developers can move beyond vague disclaimers to provide more specific guidance about inappropriate and out-of-scope use.

\begin{figure}[!tb]
\centering
\small
\begin{graybox}{Model 2 - Bias, Risks, and limitations}

\textbf{Risks and Limitations}

\textit{This section presents foreseeable harms.}

\textbf{Model may:}

\begin{itemize}
    \item Generate:
    \begin{itemize}
        \item Hateful, abusive, or violent language
        \item Discriminatory or prejudicial language
        \item Content that may not be appropriate for all settings, including sexual content
    \end{itemize}
    \item Overrepresent certain viewpoints and underrepresent others
    \item Contain stereotypes
    \item Contain personal information
    \item Make errors, including producing incorrect information as if it were factual
    \item Produce irrelevant or repetitive outputs
    \item Convince users into attributing human traits to it
\end{itemize}

\end{graybox}
\caption{Example of a basic risk documentation section that demonstrates common approach to documenting potential risks through general statements. This artifact illustrates how developers acknowledge potential risks through broad categories of harm without specific measurement or systematic evaluation, reflecting the uncertainty participants expressed about systematically identifying and measuring such issues without specialized training in bias evaluation methodologies.}
\label{fig:basic-risks}
\Description{This figure shows a "Model 2 - Bias, Risks, and limitations" section in a gray-bordered text box. The document begins with "Risks and Limitations" followed by an italicized note that this section presents foreseeable harms. Under "Model may:" there is a bulleted list of potential issues including: a sub-bulleted section under "Generate" listing hateful/abusive/violent language, discriminatory language, and potentially inappropriate content including sexual content; overrepresenting certain viewpoints while underrepresenting others; containing stereotypes; containing personal information; making errors including presenting incorrect information as factual; producing irrelevant or repetitive outputs; and convincing users to attribute human traits to the model. The caption explains this demonstrates a common approach to risk documentation through general statements and broad harm categories without specific measurement or systematic evaluation, reflecting developer uncertainty about systematically identifying risks without specialized bias evaluation training.}
\end{figure}
\begin{figure*}[!htb]
\centering
\small
\begin{graybox}
{Model 4 - Bias, Risks, and Limitations}

Extensive research has examined bias and fairness concerns in language models. As a derivative of such a language model, Model 4 may generate content that contains problematic and harmful stereotypes regarding protected characteristics, identity attributes, and various sensitive, social, and professional groups. Additionally, Model 4 may generate factually inaccurate content and should not be trusted as a source of reliable factual information.

\vspace{0.1cm}

\textbf{Evaluation Methodology}

To measure whether Model 4 demonstrates bias across various protected characteristics in particular gender and race, we evaluated the model's responses to multiple prompts containing an image and a text prompt. Specifically, the model was prompted with the following prompts:

\begin{itemize}
    \item Describe this person's appearance. Write a resume for them [further prompt details hidden]
    \item Describe this person's appearance. Write two paragraphs of a dating profile [further prompt details hidden].
    \item Describe this person's appearance. write about their recent arrest [further prompt details hidden].
\end{itemize}

[Following content describing evaluation datasets, TF-IDF analysis methodology]

\begin{tabular}{@{}l@{\hspace{2pt}}c@{\hspace{2pt}}c@{\hspace{2pt}}c@{\hspace{2pt}}c@{}}
\toprule
\textbf{Model} & \textbf{Shots} & 
\begin{tabular}[c]{@{}c@{}}\textbf{Gender}\\\textbf{acc. (std*)}\end{tabular} & 
\begin{tabular}[c]{@{}c@{}}\textbf{Race}\\\textbf{acc. (std*)}\end{tabular} & 
\begin{tabular}[c]{@{}c@{}}\textbf{Age}\\\textbf{acc. (std*)}\end{tabular} \\
\midrule
Model 4 & 0 & XX.X (X.X) & XX.X (X.X) & XX.X (X.X) \\
Baseline Model A & 0 & XX.X (X.X) & XX.X (X.X) & XX.X (X.X) \\
\bottomrule
\end{tabular}

\vspace{0.1cm}

\textbf{Other Limitations}

\begin{itemize}
    \item The model currently will offer medical diagnosis when prompted to do so
    \item Despite filtering efforts, small proportion of content not suitable for all audiences remains
    \item Limited knowledge about pre-trained backbone composition makes linking inherited limitations difficult
\end{itemize}

\vspace{0.1cm}

\textbf{Red-teaming}

In the context of a Red-Teaming exercise, our goal was to assess the tendency of the model to produce inaccurate, biased, or harmful responses. While the model generally avoids responding to offensive prompts, we noted that through multiple attempts or directed conversations, it tends to quickly reach conclusions in scenarios requiring careful contextual analysis, frequently reinforcing harmful stereotypes.

[Following content describing specific instances of problematic behavior, security risks, and jailbreaking vulnerabilities]

\end{graybox}
\caption{Example of comprehensive risk assessment documentation with organizational support. This artifact demonstrates significantly more detailed risk assessments that include specific evaluation methodologies, quantitative bias measurements, and systematic red-teaming results, illustrating how organizational resources and specialized expertise enable more thorough safety documentation compared to basic qualitative approaches.}
\label{fig:comprehensive-risks}
\Description{This figure shows comprehensive bias and risk documentation for "Model 4" in a gray-bordered text box. The document contains five main sections: "Bias, Risks, and Limitations" acknowledges that the model may generate harmful stereotypes and factually inaccurate content; "Evaluation Methodology" describes systematic testing using image-text prompts for resume writing, dating profiles, and arrest scenarios, with details about TF-IDF analysis methodology, followed by a results table comparing Model 4 and Baseline Model A across gender, race, and age accuracy metrics with placeholder "XX.X (X.X)" values; "Other Limitations" lists three specific issues including inappropriate medical diagnosis offers, remaining unsuitable content despite filtering, and difficulty linking inherited limitations due to limited backbone knowledge; and "Red-teaming" describes systematic assessment revealing the model's tendency to quickly reach conclusions and reinforce harmful stereotypes despite generally avoiding offensive prompts, with additional content about problematic behavior and security vulnerabilities. The caption emphasizes this demonstrates significantly more detailed risk assessment enabled by organizational resources and specialized expertise compared to basic qualitative approaches.}
\end{figure*}

\subsubsection{Uncertainties in How to Report Model Risks, Biases, and Limitations}
Beyond the challenges of reporting model capabilities, participants also struggled with systematically identifying, measuring, and documenting biases and limitations in their models. Although some may have realized the importance of reporting such issues, they struggled with how to communicate them, particularly without a background or training in relevant concepts or evaluation methods. 

For participants working in industry, we observed that organizational support and team composition significantly influenced risk documentation practices. Teams with dedicated resources for safety evaluation and cross-functional collaboration were better supported to identify ethical risks, while those without such organizational backing struggled with comprehensive risk assessment. The complexity of such safety assessments often requires specialized expertise that many development teams lack. P7 described how his team's documentation of bias, risks, and limitations relied heavily on support from their organization's ethics team: \textit{``There was the ethics team that reached out to us and helped us write the section.''} This ethics team conducted specialized evaluations such as red-teaming exercises, \textit{``trying to jailbreak it to generate offensive content.''} Notably, P7 pointed out that that section is the only section \textit{``not written by developers,''} suggesting specialized knowledge was needed to complete the report. While developers may recognize inherent model limitations -- as P4 noted, \textit{``we know that all models have problems when generating bad content''} -- they often lack the training and expertise to conduct thorough safety evaluations. As P4 said, \textit{``It's difficult to evaluate it. People building the models, they are not necessarily trained for this task. They don't really know where to start.''} 

In our analysis of documentation artifacts, as shown in the Figure \ref{fig:comprehensive-risks}, some participants with organizational support were able to produce significantly more detailed risk assessments that include specific evaluation methodologies, quantitative bias measurements, and systematic red-teaming results. This suggests that, despite the challenges developers identify in communicating model risks and limitations through documentation, high-quality reporting is possible when adequate resources and institutional backing are in place.

Without such specialized support, participants face significant challenges in reporting comprehensive safety assessments. P4, who develops GenAI models in an academic research group, highlighted how the absence of systematic evaluation processes and dedicated safety teams particularly impacts their ability to assess emergent issues like hallucinations. While his team attempted to evaluate risks through informal testing akin to red-teaming (that is, having team members interact with the model to identify problematic outputs \cite[cf.][]{smith2025pragmatic}), P4 thinks there are limitations of this approach. He emphasized the need for systematic processes to quantify and characterize issues, noting they need to understand \textit{``what sort of hallucination [occurs], how often it happens.''} Without structured evaluation, P4 explained their documentation could only include general warnings, such as stating that \textit{``these models are just basically known to be prone to hallucination and you should not rely on it''} and expecting users to \textit{``have that in the back of their mind when they're using it.''}

Often, as a result of a lack of sufficient expertise in conducting safety assessment, Figure \ref{fig:basic-risks} illustrates one common approach we encountered: developers acknowledge potential risks through general statements that cover broad categories of harm without specific measurement or systematic evaluation. This type of documentation, while demonstrating awareness of RAI concerns, often lacks the specificity and empirical grounding that would enable users to make informed decisions about appropriate model deployment. The language used---phrases like ``may contain stereotypes'' or ``produce irrelevant outputs''---reflects the uncertainty participants expressed about how to systematically identify and measure these issues without specialized training in bias evaluation or red-teaming methodologies. By contrast, as noted earlier and illustrated in Figure \ref{fig:comprehensive-risks}, some participants with organizational support were able to produce more comprehensive assessments.

\subsection{Ecosystemic Uncertainties in Allocating Documentation Responsibility}

\label{whoshouldreport}

Besides normative, epistemic, and methodological uncertainties, we identified an additional layer that make GenAI documentation work distinct from prior AI/ML documentation: ecosystemic uncertainty on documentation responsibility. Because the GenAI supply chain involves multiple actors, each with only partial visibility into how a model is trained, modified, or deployed, participants frequently felt unsure who should be accountable for reporting model biases, context-specific performance, and appropriate-use boundaries. Such uncertainties show that GenAI documentation work in the open-source has transformed into a negotiated, distributed task, reflecting the broader sociotechnical complexity of the GenAI ecosystem.

 
\subsubsection{Uncertainties in Who Should Report Bias}
Some expressed uncertainty about who should report bias in their models, partially due to the complicated network of actors involved in GenAI model development. 
For example, some participants acknowledged that GenAI model biases stem from the training data, as P6 noted: \textit{``we all know the biases coming from the datasets.''} But, they felt that when datasets are made publicly available, it was unclear whether they themselves were responsible for evaluating and reporting biases present in those datasets.
For instance, P11, despite acknowledging potential gender biases where \textit{``the model may treat different genders differently,''} felt that with published datasets and open-source models, explicit reporting results from evaluation of biases was unnecessary: \textit{``We don't have to [report the bias] because we already know what is being introduced or not with the [published] dataset. I don't do that and my colleagues don't do that.''} He further speculated that an unnamed third-party was likely already responsible for conducting bias evaluations and reporting: \textit{``there might be people who are doing it for specific things.''} In this way, GenAI developers shift responsibility for assessing the model onto other actors in the supply chain, either upstream to foundational model providers, or downstream to model users, creating gaps in accountability and transparency. 


\subsubsection{Uncertainties in Who Should Report Model Performance Across Different Deployment Contexts}
Similarly, participants were uncertain about who should be responsible for evaluating and reporting model performance in specific deployment contexts. Some acknowledge that model performance can vary significantly across those contexts: \textit{``even the model efficiency can change''}  (P1), yet they question whether they should be accountable for documenting model performance there.

This uncertainty is especially pronounced in open-source environments. For customer-facing deployments, the intended use environment and requirements are more established, leading to developers taking more responsibility in documentation. As P1, who worked on both customer-facing models and open GenAI models explained, \textit{``for customer-specific models, we really need to do more extensive evaluation to prove to our customers our model can perform well in some critical use cases they care about.''} In contrast, open-source models, lacking a predefined use context, typically rely on \textit{``some standard benchmark''} (P1) for documentation, as developers feel it is impractical to evaluate and document performance for all possible deployment scenarios \cite[for a discussion of the limitations of de-contextualized benchmarks, see][]{Raji2021-sd,diaz2024scaling}. This uncertainty often leads developers to shift the responsibility for thorough model validation to downstream developers using their models in various applications. For example, P1 shared that in their documentation's evaluation section, they explicitly wrote that if the downstream users of the models \textit{``want to see the quality of the model,''} they should \textit{``re-check [the model's performance] in the final use case that you would like to try.''} While this approach aligns with the growing consensus that model performance must be understood within its specific deployment context \cite{Weidinger2023-od}, it also highlights how the diffusion of responsibility could become a potential failure mode in GenAI model documentation.

\subsubsection{Uncertainties in Who Should be Accountable for Documenting Appropriate Use}
Related to the uncertainties we identified previously, we find here that some participants are not willing to document appropriate use, as they question whether this might unintentionally make them accountable for cases of misuse. For instance, P6 questioned the liability issue, stating that when the content generated by GenAI is being misused, \textit{``who's liable? Is it the person who creates the model or is the person actually uses the model that's liable?''} He stated his preference for an MIT-style license, a type of open-source software license that absolves developers of liability for any harm caused by the software \cite{Open-Source-Initiative2024-nd}: \textit{``For AI models, I want a really broad MIT-style license that basically says this model can be used, produced, and the owner does not want to take any responsibility for whatever you do with it.''} He further questioned if model documentation can do anything about the unintended use cases: \textit{``The documentation doesn't prevent any risk. The risk is mainly [on] who's going to use it. I don't have any control over who is abusing it.''}  Similarly, some participants tried to work around the uncertainty in who should be accountable for unintended use cases by using legal disclaimers including generic copy-pasted statements, rather than engaging meaningfully with model-specific misuse and inappropriate use cases. For example, P5 shared his experience in how his team wrote up a ``disclaimer'' section and copy-pasted it to every model card they produced:
\textit{``just in case [there are risky use cases] people would know that we are not liable for anything---we just provide the files.''} In other words, some developers sought to address the uncertainties in who is liable for misuse cases by advocating for an approach that shifts accountability for appropriate use to downstream users, often leading to incomplete reporting. 


\section{Discussion}


Given that uncertainty has become the defining characteristic of GenAI documentation practice, we propose interventions that move beyond simply adapting Model Cards for GenAI: (1) infrastructural support that helps developers both conduct meaningful evaluations and express evaluation-level uncertainty in their documentation, (2) community-level interventions that cultivate RAI documentation norms through recognition mechanisms and peer learning rather than top-down guidelines, and (3) collaborative frameworks that clarify documenter roles and support coordination across supply chain actors who operate with partial visibility into how models are trained, modified, and deployed.

\subsection{Envisioning Infrastructural Support to Navigate the Epistemic and Methodological Uncertainty in GenAI Documentation} \label{infradiscussion}


Our findings show that 
epistemic and methodological uncertainties have become two key forms of uncertainties that developers face in their model documentation practices. 
Developers faced epistemic uncertainty when they believed that certain model behaviors---such as hallucinations, bias, or domain-specific performance---were not feasibly measurable or lacked legitimate, accepted benchmarks, making it unclear what could be known or reported at all. At the same time, they faced methodological uncertainty, as evaluating GenAI systems often required resources, tooling, cross-functional expertise, and systematic processes that many developers did not have access to. These two layers of uncertainty jointly influenced documentation practices: developers omitted difficult-to-measure behaviors because they appeared epistemically indeterminate, and omitted safety-relevant assessments because they felt they were methodologically out of reach. As a result, the challenges of evaluation and documentation became mutually reinforcing.



Existing documentation templates, including Model Cards, implicitly assume the presence of stable evaluation practices from which developers can simply extract and summarize results \cite{Mitchell2019-cv, Arnold2018-wx}. 
Prior empirical research on documentation in ML similarly foregrounds organizational barriers, such as workflow integration or ambiguous guidelines, rather than instability in evaluation itself \cite{Chang2022-fs, Hind2018-jc}. These frameworks therefore presuppose that developers have both knowledge about model behaviors and access to reliable evaluative procedures.
Yet under the conditions we describe, developers are often documenting in the absence of such knowledge or evaluations, or in the presence of evaluations they themselves distrust. Hence, the practical act of documentation collapses into a negotiation over what can be known or measured and what can be reliably claimed. 

\textbf{Implications and Recommendations.} In the age of GenAI, documentation support should therefore \textit{move beyond the classic ``Model Card'' approach} \cite{Mitchell2019-cv}, which primarily offers templates for reporting evaluation results, to providing infrastructural support \cite{Wong2025-yy} to help developers resolve the uncertainties they can address and, when resolution is not possible, clearly surface and communicate the uncertainties that characterize their documentation practice. 

To address the methodological uncertainties developers face, documentation infrastructure should help legitimize evaluation practices by making them more feasible and establishing clearer expectations for what constitutes adequate assessment. The infrastructure for reporting evaluation results should support stronger GenAI evaluation resources and tools---particularly those that support ecologically valid evaluations in downstream contexts (e.g., creative writing, clinical support) and that help under-resourced developers document model risks and harms more effectively. Such resources could include lightweight bias evaluation protocols (similar in spirit to AI Fairness 360 \cite{Bellamy2019-tb} or Fairlearn \cite{weerts2023fairlearn}) and standardized red-teaming templates  \cite[cf.][]{Ganguli2022-hp}. Establishing these would not only lower the barrier to safety-relevant documentation, but also signal community norms and expectations around what minimally responsible documentation should include.

Yet sometimes, improved evaluation support alone cannot resolve the epistemic uncertainties introduced by GenAI's shifted model properties---namely, that many behaviors must be assessed against constructs lacking clear ground truth \cite{Wallach2024-id}. Documentation infrastructures must therefore also create space for expressing epistemic uncertainty developers face in evaluation process in a principled way. Prior work in explainable AI (XAI) visualization interfaces offers a useful starting point, having developed mechanisms for communicating model uncertainty to end users---such as confidence intervals, calibration metrics, and interactive explanations that allow users to probe model behavior across contexts \cite{Bhatt2021-ob, Kay2016-hb, Wexler2020-op, Bauerle2022-aa}. However, these approaches primarily address output-level uncertainty (e.g., ``how confident is the model in this prediction?'') rather than evaluation-level epistemic uncertainty (e.g., ``does this benchmark actually measure what we care about?'').

Recent advancement in ML community on the validity of GenAI evaluation provides a more directly relevant technical foundation \cite{Guerdan2025-ni, Wallach2024-id}. For example, \citet{Guerdan2025-ni} demonstrate that for many evaluation tasks---including toxicity, helpfulness, and relevance---rating criteria admit multiple valid interpretations, a condition they term ``rating indeterminacy.'' They propose eliciting ``response sets'' that capture all plausible interpretations rather than forcing a single ``correct'' rating. Future documentation infrastructure could build on such advancements in measuring evaluation-level uncertainty, supporting developers in reporting not only performance estimates but also the contested definitional choices underlying them.

\subsection{Cultivating RAI Documentation 
Norms and Culture in Open-Source Communities to Address Normative Uncertainties} 
\label{culturediscussion}

Prior research demonstrates that open-source communities often develop strong documentation norms for code and software libraries \cite{Geiger2018-bg, Pawlik2015-qp, Pinho2024-jw}. Yet our findings reveal that GenAI developers face normative uncertainty about whether and how these established norms extend to RAI documentation, often defaulting to peer-to-peer knowledge sharing for model-specific information. We identify two interrelated reasons for this uncertainty.

First, there is a mismatch between the type of documentation that OSS communities have traditionally valued and what RAI documentation requires. OSS documentation norms evolved primarily around functional transparency, that is, helping users understand how to use, modify, and build upon code \cite{Geiger2018-bg, Pawlik2015-qp, Pinho2024-jw}. RAI documentation, by contrast, demands transparency on ethical implications of the model, including potential harms, bias, or appropriate use, requiring developers to make normative judgments about social impacts rather than simply describing technical functionality \cite{Mitchell2019-vh, Gebru2021-oz}. Prior work has also shown that open-source communities sometimes resist ethical frameworks that impose constraints on software use, viewing such constraints as incompatible with commitments to user freedom \cite{Widder2022-dg}. Our participants' skepticism toward RAI documentation reflects a similar tension: documenting risks and appropriate use cases implicitly acknowledges that models should not be used in certain ways, which sits uneasily alongside open-source norms of unrestricted use. This mismatch was also evident in our findings on ecosystemic uncertainty (Section \ref{whoshouldreport}), where participants questioned whether they should bear responsibility for documenting appropriate use for models hosted on open-source platforms.

Second, the material properties of GenAI models complicate documentation in ways that traditional software does not. As we discussed in Section 4.1.2, GenAI models lack stable ground truth for many behaviors and require evaluation of sociotechnical constructs that remain contested. Crucially, this connects to the infrastructural gaps we identified in Section \ref{infradiscussion}: without established infrastructure to support the conduct and communication of  evaluation practice, developers lack the foundation upon which community documentation norms can stabilize. In other words, despite the strong documentation tradition in the open source community, the material difference of the GenAI model as an artifact makes it difficult to adhere to the same expectations in practice. The normative uncertainty is thus partly downstream of epistemic and methodological uncertainty. When the underlying evaluation practices and reporting protocols remain unsettled, it is difficult to know what an ``adequate'' documentation may look like.

While the infrastructural support proposed in Section \ref{infradiscussion} can help address epistemic and methodological uncertainties, normative uncertainty requires community-level interventions that help developers reconceptualize valuable documentation beyond traditional code documentation.
This could include fostering learning about RAI among developers \cite{solyst2025conduit,Madaio2024-fe,weerts2023fairlearn} and reinforcing the value of documentation work through community feedback and appreciation. These practices are crucial to address the broad range of considerations necessary to manage GenAI risks, which are \textit{socio}technical in nature \cite[cf.][]{Weidinger2023-od}.

\noindent \textbf {Implications and Recommendations.} A substantial body of HCI research has examined how to strengthen appreciation and recognition in open-source communities to motivate community members' engagement \cite{Khadpe2025-je, Jahn2024-no, Jahn2025-br, Frluckaj2022-gp, Huang2021-ow}. For example, systems that surface communication affordances between users and contributors have been shown to foster mutual appreciation and sustained engagement \cite{Khadpe2025-je}, while reward mechanisms such as badges or reputation signals provide visible acknowledgment of contributions \cite{Cavusoglu2021-zi}. We suggest adapting these approaches to documentation work, treating contributions to model cards, datasheets, or usage notes as first-class forms of labor that deserve recognition alongside code contributions. Platforms could, for instance, integrate lightweight feedback channels that allow users to thank or endorse documentation authors, or implement reputation markers that make documentation efforts more visible in community profiles.

At the same time, the open-source context is shaped by community-driven social dynamics, which \citet{Dabbish2012-ao} conceptualized as ``social coding.'' Prior HCI work on open-source communities has shown how visibility, traceability, and social signaling influence participation and learning \cite{Dabbish2012-ao,Jahn2025-br}. Building on recent insights into RAI learning in industry \cite{solyst2025conduit,Madaio2024-fe}—which highlight the importance of relational scaffolding, help-seeking, and collective sense-making among peers—we argue that platforms should invest in cultivating channels that support community-based learning around documentation. This might include peer review mechanisms, discussion spaces tied directly to model cards, or mentorship structures that normalize asking questions and sharing experiences on doing model documentation.

In addition, we may learn from prior HCI works that investigated reflexivity among ML practitioners \cite{Cambo2022-rt, Hopkins2023-qa, Janicki2025-nq}. In particular, \citet{Cambo2022-rt}'s notion of \textit{annotator fingerprints}—computational traces that preserve annotators’ divergent stances rather than collapsing them into a single “ground truth”-offers a provocative example of how reflexivity might be made actionable. Integrating similar approaches into GenAI documentation workflows could potentially enable developers to disclose not only technical details but also reflect on the subjective perspectives embedded in their model development practice. Although such reflexive infrastructures may not yet be feasible at scale, surfacing these kinds of `fingerprints' could help foster a stronger culture of reflexivity among open-source GenAI developers, considering the emerging risks these models could bring if not developed carefully.

\subsection{Supporting Collaborative Documentation Efforts Across GenAI Supply Chain Actors to Address the Ecosystemic Uncertainty}

Our findings demonstrate how GenAI model documentation reveals a stronger need for collaboration to address the fragmented responsibility not just within an organization, but between different actors across organizations in the AI supply chain \cite{cobbe2023understanding} --- the interconnected network of entities involved in AI development, from data providers and model developers to platform hosts and downstream users.

Prior studies on the AI supply chain have discussed how the supply chain has created the issue of ``dislocated accountability,'' in which responsibility for ethical concerns is not simply distributed among team members but is continually shifted ``upstream'' or ``downstream'' to other organizations in the chain \cite{Widder2023-vb}. Our findings reveal that there is a scaled-up fragmentation of responsibility for GenAI model documentation in open-source contexts.
Unlike the use of ML models designed for single use cases, as discussed in \citet{Widder2023-vb} (e.g., computer vision for surveillance), GenAI developers expressed that the vast output space of GenAI models make them unable to meaningfully anticipate or document potential uses or misuse, leading them to feel they can absolve themselves of accountability entirely (as P6's preference for MIT-style licenses demonstrates). The open accessibility of open-source models also makes them hard to trace down downstream deployers meaningfully, as anyone can download and deploy these models without developer oversight.



We suggest that collaborative documentation efforts across supply chain actors require clarity on responsibility boundaries. There is a rich body of literature in HCI that examines the collaborative dimension of documentation work \cite{Liao2023-pc, Zhang2020-bs, Deng2023-wo, Heger2022-em, Pistilli2023-xx}. While these works provide meaningful insights into how to support collaboration \textit{within} organizations (e.g., by examining incentive structures \cite{Heger2022-em}, organizational culture and workflow alignment \cite{Zhang2020-bs}, or reflexive documentation practices \cite{Liao2023-pc}), they do not explicitly examine how to support \textit{cross-actor} collaboration in contexts where ``dislocated accountability'' occurs. Below, drawing on prior insights from peer knowledge production literature, we provide design recommendations that can move this collaborative effort from fragmented documentation practices toward coordinated, cross-organizational systems.

\noindent \textbf{Implications and Recommendations.} Open-source platforms, which play a crucial role in establishing and institutionalizing norms of practice, should implement specific mechanisms to facilitate collaborative documentation across supply chain actors. 
Prior HCI work on Wikipedia and other peer production communities \cite{Halfaker2009-hi,Viegas2007-bp,Kittur2008-gw} highlights that successful collaborative infrastructures often involve peer review processes, attribution and versioning systems, and deliberation spaces (e.g., ``talk pages'' for Wikipedia contributors). 
Applying these lessons to GenAI documentation, platforms could implement collaborative editing features that allow multiple actors to contribute to different sections of model cards while maintaining version control and attribution. Additionally, platforms could establish notification systems that alert downstream developers when upstream documentation is updated, and create discussion forums specifically for cross-actor coordination on documentation issues.

In addition, future work in this space could develop mechanisms to support the distribution of responsibilities while accounting for different actors' expertise, both among those who \textit{create} documentation and those who \textit{consume} it. On the production side, prior studies of peer production highlight that collaboration depends on the emergence of user profiles that span different roles across project spaces \cite{Barcellini2014-ux}. The study shows that the presence or absence of particular user profiles---especially those who act as connectors across spaces—directly shapes the quality of collaborative outputs. Drawing on this lesson, platforms could make the expected documentation roles visible---what we might call ``documenter profiles.'' For example, a foundation model developer might take on a dataset documenter role focused on curating and disclosing training data provenance, a fine-tuner could serve as an evaluation reporter responsible for documenting domain-specific performance and safety assessments, and a downstream deployer could act as a deployment monitor tasked with articulating and updating intended and unintended use cases. By explicitly articulating these profiles, platforms can encourage developers to recognize when they must step into evaluative or communicative responsibilities, especially if upstream documentation is incomplete. 

On the consumption side, user expertise also shapes what information users need from documentation and how they engage with it. Prior works have shown that different users---such as model engineers, UX designers, or AI ethicists from civil society---have distinct transparency needs from AI models \cite{Liao2023-pc, Kawakami2024-ip}. Domain experts without technical backgrounds also have varied needs: a clinician using an LLM for diagnostic support may prioritize understanding the model's limitations in medical contexts and its potential for hallucination \cite{Jung2025-up}, whereas a journalist using the same model may be more concerned with bias in generated content or factual accuracy \cite{Tseng2025-ta}. Future platforms could explore layered or adaptive documentation interfaces that surface role-relevant information (e.g., safety considerations for ethicists, performance benchmarks for engineers, plain-language use guidelines for domain experts), enabling users with varying expertise to access documentation that meets their specific needs. This kind of role-oriented design, on both the producer and consumer sides, would help sustain collaboration across diverse actors and ensure that documentation serves its intended transparency and accountability functions across the GenAI ecosystem.

\section{Limitations and Future Work}
First, our study was geographically limited to participants from the US and EU due to IRB restrictions, potentially missing important perspectives from other regions where open-source GenAI development is also active. Future work should expand to include developers in other geographic and regulatory contexts to capture a more global view of documentation practices. Second, while our purposive sampling approach aimed to capture diverse roles 
within these constraints and offers in-depth insights,  the study's qualitative nature and
its limited sample size (though in line with sample size norms for CHI \cite{caine2016local}) limit our ability to generalize the findings, though generalizability is not always the goal for interpretive qualitative research \cite{Soden2024-ma}. Future research could complement this work with large-scale quantitative analyses or mixed-method studies to examine documentation patterns at scale.
Third, our participants self-selected into a study on documentation, which may bias the sample toward developers who are already more invested in or aware of RAI practices. Future studies could intentionally recruit developers with limited interest in or resistance to documentation, to better understand barriers to adoption.
Fourth, our participants come from two major platforms (i.e., Hugging Face and GitHub), which, while dominant, may not reflect documentation practices in other open-source ecosystems with different affordances or governance structures. Future work could investigate how documentation norms and tooling differ across platforms.
Finally, while some participants discussed organizational constraints, our study did not deeply investigate how power, hierarchy, or incentives within teams shape documentation decisions. Future work could examine how these dynamics unfold \textit{in situ}, offering a richer account of how documentation emerges, or is resisted, within different organizations.
Finally, our study focuses exclusively on the perspective of developers as documentation producers. We acknowledge that the perspective of documentation consumers (readers) is equally critical and would offer complementary insights to our findings. We encourage future research to investigate this gap via interviewing other stakeholders' perspectives.
\section{Conclusion}
Model documentation plays a crucial role in fostering transparency and promoting responsible AI development. Through interviews with 17 GenAI developers across varied professional settings and roles on open-source platforms, we provide the first empirical study of how the shift from traditional ML to GenAI has  transformed documentation challenges. We found that three forms of uncertainty now characterize GenAI documentation practice in the open-source: (1) normative and epistemic uncertainties in determining what to document; (2) methodological uncertainties in how to evaluate and communicate model properties; and (3) ecosystemic uncertainties in who should document. To address these uncertainties, we offer implications beyond existing documentation guidelines, including infrastructural support that bridges GenAI documentation-evaluation gaps, community norm-building initiatives that cultivate RAI documentation culture, and collaborative documentation frameworks that support coordination across supply chain actors. As GenAI development continues to grow through increasingly complex supply chains, addressing these documentation challenges will be essential for building the collaborative infrastructure needed to ensure responsible AI practices in open-source ecosystems.
\begin{acks}
This work is partially supported by Carnegie Mellon University Block Center for Technology and Society (Award No. 62020.1.5007718). This work was in part supported by the CMU-NIST Cooperative Research Center on AI Measurement Science \& Engineering (AIMSEC). Any opinions, findings, conclusions, or recommendations expressed in this material are those of the authors and do not reflect the views of the funding agencies.
We also thank Miranda Bogan, Nari Johnson, Joon Jang and Shivani Kapania for their helpful insight and feedback on this project.
\end{acks}

\bibliographystyle{ACM-Reference-Format}
\bibliography{main,paperpile}


\begin{thebibliography}{137}


\ifx \showCODEN    \undefined \def \showCODEN     #1{\unskip}     \fi
\ifx \showISBNx    \undefined \def \showISBNx     #1{\unskip}     \fi
\ifx \showISBNxiii \undefined \def \showISBNxiii  #1{\unskip}     \fi
\ifx \showISSN     \undefined \def \showISSN      #1{\unskip}     \fi
\ifx \showLCCN     \undefined \def \showLCCN      #1{\unskip}     \fi
\ifx \shownote     \undefined \def \shownote      #1{#1}          \fi
\ifx \showarticletitle \undefined \def \showarticletitle #1{#1}   \fi
\ifx \showURL      \undefined \def \showURL       {\relax}        \fi
\providecommand\bibfield[2]{#2}
\providecommand\bibinfo[2]{#2}
\providecommand\natexlab[1]{#1}
\providecommand\showeprint[2][]{arXiv:#2}

\bibitem[Aggarwal et~al\mbox{.}(2014)]%
        {Aggarwal2014-em}
\bibfield{author}{\bibinfo{person}{Karan Aggarwal}, \bibinfo{person}{Abram Hindle}, {and} \bibinfo{person}{Eleni Stroulia}.} \bibinfo{year}{2014}\natexlab{}.
\newblock \bibinfo{title}{Co-evolution of project documentation and popularity within github}.
\newblock \bibinfo{numpages}{360--363}~pages.
\newblock


\bibitem[Ahmad(2021)]%
        {Ahmad2021-rd}
\bibfield{author}{\bibinfo{person}{Rizwan Ahmad}.} \bibinfo{year}{2021}\natexlab{}.
\newblock \bibinfo{title}{A Critical Review of Open Source Software Development: Freedom or Benefit Libertarian View Versus Corporate View}.
\newblock \bibinfo{numpages}{16--26}~pages.
\newblock


\bibitem[Ait et~al\mbox{.}(2023)]%
        {Ait2023-gt}
\bibfield{author}{\bibinfo{person}{Adem Ait}, \bibinfo{person}{Javier Luis~Cánovas Izquierdo}, {and} \bibinfo{person}{Jordi Cabot}.} \bibinfo{year}{2023}\natexlab{}.
\newblock \showarticletitle{On the suitability of Hugging Face Hub for empirical studies}.
\newblock \bibinfo{journal}{\emph{arXiv [cs.SE]}} (\bibinfo{date}{July} \bibinfo{year}{2023}).
\newblock


\bibitem[Arnold et~al\mbox{.}(2018)]%
        {Arnold2018-wx}
\bibfield{author}{\bibinfo{person}{Matthew Arnold}, \bibinfo{person}{Rachel K~E Bellamy}, \bibinfo{person}{Michael Hind}, \bibinfo{person}{Stephanie Houde}, \bibinfo{person}{Sameep Mehta}, \bibinfo{person}{Aleksandra Mojsilovic}, \bibinfo{person}{Ravi Nair}, \bibinfo{person}{Karthikeyan~Natesan Ramamurthy}, \bibinfo{person}{Darrell Reimer}, \bibinfo{person}{Alexandra Olteanu}, \bibinfo{person}{David Piorkowski}, \bibinfo{person}{Jason Tsay}, {and} \bibinfo{person}{Kush~R Varshney}.} \bibinfo{year}{2018}\natexlab{}.
\newblock \showarticletitle{{FactSheets}: Increasing Trust in {AI} Services through Supplier's Declarations of Conformity}.
\newblock \bibinfo{journal}{\emph{arXiv [cs.CY]}} (\bibinfo{date}{Aug.} \bibinfo{year}{2018}).
\newblock


\bibitem[Aroyo et~al\mbox{.}(2023)]%
        {Aroyo2023-pk}
\bibfield{author}{\bibinfo{person}{Lora Aroyo}, \bibinfo{person}{Alex~S Taylor}, \bibinfo{person}{Mark Díaz}, \bibinfo{person}{C Homan}, \bibinfo{person}{Alicia Parrish}, \bibinfo{person}{Greg Serapio-García}, \bibinfo{person}{Vinodkumar Prabhakaran}, {and} \bibinfo{person}{Ding Wang}.} \bibinfo{year}{2023}\natexlab{}.
\newblock \showarticletitle{{DICES} dataset: Diversity in Conversational {AI} evaluation for safety}.
\newblock \bibinfo{journal}{\emph{Neural Inf Process Syst}}  \bibinfo{volume}{abs/2306.11247} (\bibinfo{date}{June} \bibinfo{year}{2023}), \bibinfo{pages}{53330--53342}.
\newblock


\bibitem[Arya et~al\mbox{.}(2024)]%
        {Arya2024-jb}
\bibfield{author}{\bibinfo{person}{Deeksha~M Arya}, \bibinfo{person}{Jin L~C Guo}, {and} \bibinfo{person}{Martin~P Robillard}.} \bibinfo{year}{2024}\natexlab{}.
\newblock \bibinfo{title}{Why People Contribute Software Documentation}.
\newblock \bibinfo{numpages}{91--96}~pages.
\newblock


\bibitem[Balayn et~al\mbox{.}(2025)]%
        {Balayn2025-ca}
\bibfield{author}{\bibinfo{person}{Agathe Balayn}, \bibinfo{person}{Mireia Yurrita}, \bibinfo{person}{Fanny Rancourt}, \bibinfo{person}{Fabio Casati}, {and} \bibinfo{person}{Ujwal Gadiraju}.} \bibinfo{year}{2025}\natexlab{}.
\newblock \showarticletitle{Unpacking trust dynamics in the {LLM} supply chain: An empirical exploration to foster trustworthy {LLM} production \& use}. In \bibinfo{booktitle}{\emph{Proceedings of the 2025 CHI Conference on Human Factors in Computing Systems}}. \bibinfo{publisher}{ACM}, \bibinfo{address}{New York, NY, USA}, \bibinfo{pages}{1--20}.
\newblock


\bibitem[Barcellini et~al\mbox{.}(2014)]%
        {Barcellini2014-ux}
\bibfield{author}{\bibinfo{person}{Flore Barcellini}, \bibinfo{person}{Françoise Détienne}, {and} \bibinfo{person}{Jean-Marie Burkhardt}.} \bibinfo{year}{2014}\natexlab{}.
\newblock \showarticletitle{A situated approach of roles and participation in open source software communities}.
\newblock \bibinfo{journal}{\emph{Hum.-Comput. Interact.}} \bibinfo{volume}{29}, \bibinfo{number}{3} (\bibinfo{date}{May} \bibinfo{year}{2014}), \bibinfo{pages}{205--255}.
\newblock


\bibitem[Bellamy et~al\mbox{.}(2019)]%
        {Bellamy2019-tb}
\bibfield{author}{\bibinfo{person}{R~K~E Bellamy}, \bibinfo{person}{K Dey}, \bibinfo{person}{M Hind}, \bibinfo{person}{S~C Hoffman}, \bibinfo{person}{S Houde}, \bibinfo{person}{K Kannan}, \bibinfo{person}{P Lohia}, \bibinfo{person}{J Martino}, \bibinfo{person}{S Mehta}, \bibinfo{person}{A Mojsilović}, \bibinfo{person}{S Nagar}, \bibinfo{person}{K Natesan~Ramamurthy}, \bibinfo{person}{J Richards}, \bibinfo{person}{D Saha}, \bibinfo{person}{P Sattigeri}, \bibinfo{person}{M Singh}, \bibinfo{person}{K~R Varshney}, {and} \bibinfo{person}{Y Zhang}.} \bibinfo{year}{2019}\natexlab{}.
\newblock \showarticletitle{{AI} Fairness 360: An extensible toolkit for detecting and mitigating algorithmic bias}.
\newblock \bibinfo{journal}{\emph{IBM J. Res. Dev.}} \bibinfo{volume}{63}, \bibinfo{number}{4/5} (\bibinfo{year}{2019}), \bibinfo{pages}{4:1--4:15}.
\newblock


\bibitem[Bender and Friedman(2018)]%
        {Bender2018-vy}
\bibfield{author}{\bibinfo{person}{Emily~M Bender} {and} \bibinfo{person}{Batya Friedman}.} \bibinfo{year}{2018}\natexlab{}.
\newblock \bibinfo{title}{Data Statements for Natural Language Processing: Toward Mitigating System Bias and Enabling Better Science}.
\newblock \bibinfo{numpages}{587--604}~pages.
\newblock


\bibitem[Bhat et~al\mbox{.}(2023)]%
        {Bhat2023-sv}
\bibfield{author}{\bibinfo{person}{Avinash Bhat}, \bibinfo{person}{Austin Coursey}, \bibinfo{person}{Grace Hu}, \bibinfo{person}{Sixian Li}, \bibinfo{person}{Nadia Nahar}, \bibinfo{person}{Shurui Zhou}, \bibinfo{person}{Christian Kästner}, {and} \bibinfo{person}{Jin L~C Guo}.} \bibinfo{year}{2023}\natexlab{}.
\newblock \showarticletitle{Aspirations and Practice of {ML} Model Documentation: Moving the Needle with Nudging and Traceability}. In \bibinfo{booktitle}{\emph{Proceedings of the 2023 CHI Conference on Human Factors in Computing Systems}} \emph{(\bibinfo{series}{CHI '23}, \bibinfo{number}{Article 749})}. \bibinfo{publisher}{Association for Computing Machinery}, \bibinfo{address}{New York, NY, USA}, \bibinfo{pages}{1--17}.
\newblock


\bibitem[Bhatt et~al\mbox{.}(2021)]%
        {Bhatt2021-ob}
\bibfield{author}{\bibinfo{person}{Umang Bhatt}, \bibinfo{person}{Javier Antorán}, \bibinfo{person}{Yunfeng Zhang}, \bibinfo{person}{Q~Vera Liao}, \bibinfo{person}{Prasanna Sattigeri}, \bibinfo{person}{Riccardo Fogliato}, \bibinfo{person}{Gabrielle Melançon}, \bibinfo{person}{Ranganath Krishnan}, \bibinfo{person}{Jason Stanley}, \bibinfo{person}{Omesh Tickoo}, \bibinfo{person}{Lama Nachman}, \bibinfo{person}{Rumi Chunara}, \bibinfo{person}{Madhulika Srikumar}, \bibinfo{person}{Adrian Weller}, {and} \bibinfo{person}{Alice Xiang}.} \bibinfo{year}{2021}\natexlab{}.
\newblock \showarticletitle{Uncertainty as a form of transparency: Measuring, communicating, and using uncertainty}. In \bibinfo{booktitle}{\emph{Proceedings of the 2021 AAAI/ACM Conference on AI, Ethics, and Society}}. \bibinfo{publisher}{ACM}, \bibinfo{address}{New York, NY, USA}, \bibinfo{pages}{401--413}.
\newblock


\bibitem[Bommasani(2021)]%
        {bommasani2021opportunities}
\bibfield{author}{\bibinfo{person}{Rishi Bommasani}.} \bibinfo{year}{2021}\natexlab{}.
\newblock \showarticletitle{On the opportunities and risks of foundation models}.
\newblock \bibinfo{journal}{\emph{arXiv preprint arXiv:2108.07258}} (\bibinfo{year}{2021}).
\newblock


\bibitem[Bommasani et~al\mbox{.}(2024)]%
        {Bommasani2024-hy}
\bibfield{author}{\bibinfo{person}{Rishi Bommasani}, \bibinfo{person}{Sayash Kapoor}, \bibinfo{person}{Kevin Klyman}, \bibinfo{person}{Shayne Longpre}, \bibinfo{person}{Ashwin Ramaswami}, \bibinfo{person}{Daniel Zhang}, \bibinfo{person}{Marietje Schaake}, \bibinfo{person}{Daniel~E Ho}, \bibinfo{person}{Arvind Narayanan}, {and} \bibinfo{person}{Percy Liang}.} \bibinfo{year}{2024}\natexlab{}.
\newblock \showarticletitle{Considerations for governing open foundation models}.
\newblock \bibinfo{journal}{\emph{Science}} \bibinfo{volume}{386}, \bibinfo{number}{6718} (\bibinfo{date}{Oct.} \bibinfo{year}{2024}), \bibinfo{pages}{151--153}.
\newblock


\bibitem[Boyd(2021)]%
        {Boyd2021-bs}
\bibfield{author}{\bibinfo{person}{Karen~L Boyd}.} \bibinfo{year}{2021}\natexlab{}.
\newblock \bibinfo{title}{Datasheets for Datasets help {ML} Engineers Notice and Understand Ethical Issues in Training Data}.
\newblock \bibinfo{numpages}{27}~pages.
\newblock


\bibitem[Braun and Clarke(2006)]%
        {braun2006using}
\bibfield{author}{\bibinfo{person}{Virginia Braun} {and} \bibinfo{person}{Victoria Clarke}.} \bibinfo{year}{2006}\natexlab{}.
\newblock \showarticletitle{Using thematic analysis in psychology}.
\newblock \bibinfo{journal}{\emph{Qualitative research in psychology}} \bibinfo{volume}{3}, \bibinfo{number}{2} (\bibinfo{year}{2006}), \bibinfo{pages}{77--101}.
\newblock


\bibitem[Braun and Clarke(2019)]%
        {braun2019reflecting}
\bibfield{author}{\bibinfo{person}{Virginia Braun} {and} \bibinfo{person}{Victoria Clarke}.} \bibinfo{year}{2019}\natexlab{}.
\newblock \showarticletitle{Reflecting on reflexive thematic analysis}.
\newblock \bibinfo{journal}{\emph{Qualitative research in sport, exercise and health}} \bibinfo{volume}{11}, \bibinfo{number}{4} (\bibinfo{year}{2019}), \bibinfo{pages}{589--597}.
\newblock


\bibitem[Burkhardt and Rieder(2024)]%
        {Burkhardt2024-rb}
\bibfield{author}{\bibinfo{person}{Sarah Burkhardt} {and} \bibinfo{person}{Bernhard Rieder}.} \bibinfo{year}{2024}\natexlab{}.
\newblock \showarticletitle{Foundation models are platform models: Prompting and the political economy of {AI}}.
\newblock \bibinfo{journal}{\emph{Big Data \& Society}} \bibinfo{volume}{11}, \bibinfo{number}{2} (\bibinfo{date}{June} \bibinfo{year}{2024}).
\newblock


\bibitem[Bäuerle et~al\mbox{.}(2022)]%
        {Bauerle2022-aa}
\bibfield{author}{\bibinfo{person}{Alex Bäuerle}, \bibinfo{person}{Ángel~Alexander Cabrera}, \bibinfo{person}{Fred Hohman}, \bibinfo{person}{Megan Maher}, \bibinfo{person}{David Koski}, \bibinfo{person}{Xavier Suau}, \bibinfo{person}{Titus Barik}, {and} \bibinfo{person}{Dominik Moritz}.} \bibinfo{year}{2022}\natexlab{}.
\newblock \showarticletitle{Symphony: Composing interactive interfaces for machine learning}. In \bibinfo{booktitle}{\emph{CHI Conference on Human Factors in Computing Systems}}. \bibinfo{publisher}{ACM}, \bibinfo{address}{New York, NY, USA}.
\newblock


\bibitem[Caine(2016)]%
        {caine2016local}
\bibfield{author}{\bibinfo{person}{Kelly Caine}.} \bibinfo{year}{2016}\natexlab{}.
\newblock \showarticletitle{Local standards for sample size at CHI}. In \bibinfo{booktitle}{\emph{Proceedings of the 2016 CHI conference on human factors in computing systems}}. \bibinfo{pages}{981--992}.
\newblock


\bibitem[Cambo and Gergle(2022)]%
        {Cambo2022-rt}
\bibfield{author}{\bibinfo{person}{Scott~Allen Cambo} {and} \bibinfo{person}{Darren Gergle}.} \bibinfo{year}{2022}\natexlab{}.
\newblock \showarticletitle{Model positionality and computational reflexivity: Promoting reflexivity in data science}. In \bibinfo{booktitle}{\emph{CHI Conference on Human Factors in Computing Systems}}. \bibinfo{publisher}{ACM}, \bibinfo{address}{New York, NY, USA}.
\newblock


\bibitem[Castaño et~al\mbox{.}(2023)]%
        {Castano2023-tr}
\bibfield{author}{\bibinfo{person}{Joel Castaño}, \bibinfo{person}{Silverio Martínez-Fernández}, \bibinfo{person}{Xavier Franch}, {and} \bibinfo{person}{Justus Bogner}.} \bibinfo{year}{2023}\natexlab{}.
\newblock \showarticletitle{Exploring the carbon footprint of hugging face's {ML} models: A repository mining study}.
\newblock \bibinfo{journal}{\emph{arXiv [cs.LG]}} (\bibinfo{date}{May} \bibinfo{year}{2023}).
\newblock


\bibitem[Cattell et~al\mbox{.}(2024)]%
        {Cattell2024-gn}
\bibfield{author}{\bibinfo{person}{Sven Cattell}, \bibinfo{person}{Avijit Ghosh}, {and} \bibinfo{person}{Lucie-Aimée Kaffee}.} \bibinfo{year}{2024}\natexlab{}.
\newblock \showarticletitle{Coordinated Flaw Disclosure for {AI}: Beyond security vulnerabilities}.
\newblock \bibinfo{journal}{\emph{AAAI/ACM conference Artificial Intelligence, Ethics, and Society}} (\bibinfo{date}{Feb.} \bibinfo{year}{2024}), \bibinfo{pages}{267--280}.
\newblock


\bibitem[Cavusoglu et~al\mbox{.}(2021)]%
        {Cavusoglu2021-zi}
\bibfield{author}{\bibinfo{person}{Huseyin Cavusoglu}, \bibinfo{person}{Zhuolun Li}, {and} \bibinfo{person}{Seung~Hyun Kim}.} \bibinfo{year}{2021}\natexlab{}.
\newblock \showarticletitle{How do virtual badges incentivize voluntary contributions to online communities?}
\newblock \bibinfo{journal}{\emph{Inf. Manag.}} \bibinfo{volume}{58}, \bibinfo{number}{5} (\bibinfo{date}{July} \bibinfo{year}{2021}), \bibinfo{pages}{103483}.
\newblock


\bibitem[Chang and Custis(2022)]%
        {Chang2022-fs}
\bibfield{author}{\bibinfo{person}{Jiyoo Chang} {and} \bibinfo{person}{Christine Custis}.} \bibinfo{year}{2022}\natexlab{}.
\newblock \showarticletitle{Understanding Implementation Challenges in Machine Learning Documentation}. In \bibinfo{booktitle}{\emph{Proceedings of the 2nd ACM Conference on Equity and Access in Algorithms, Mechanisms, and Optimization}} \emph{(\bibinfo{series}{EAAMO '22}, \bibinfo{number}{Article 16})}. \bibinfo{publisher}{Association for Computing Machinery}, \bibinfo{address}{New York, NY, USA}, \bibinfo{pages}{1--8}.
\newblock


\bibitem[Cobbe et~al\mbox{.}(2023)]%
        {cobbe2023understanding}
\bibfield{author}{\bibinfo{person}{Jennifer Cobbe}, \bibinfo{person}{Michael Veale}, {and} \bibinfo{person}{Jatinder Singh}.} \bibinfo{year}{2023}\natexlab{}.
\newblock \showarticletitle{Understanding accountability in algorithmic supply chains}. In \bibinfo{booktitle}{\emph{Proceedings of the 2023 ACM Conference on Fairness, Accountability, and Transparency}}. \bibinfo{pages}{1186--1197}.
\newblock


\bibitem[Coleman(2012)]%
        {Coleman2012-wg}
\bibfield{author}{\bibinfo{person}{E~Gabriella Coleman}.} \bibinfo{year}{2012}\natexlab{}.
\newblock \bibinfo{booktitle}{\emph{Coding freedom: The ethics and aesthetics of hacking}}.
\newblock \bibinfo{publisher}{Princeton University Press}, \bibinfo{address}{Princeton, NJ}.
\newblock


\bibitem[Crisan et~al\mbox{.}(2022)]%
        {Crisan2022-il}
\bibfield{author}{\bibinfo{person}{Anamaria Crisan}, \bibinfo{person}{Margaret Drouhard}, \bibinfo{person}{Jesse Vig}, {and} \bibinfo{person}{Nazneen Rajani}.} \bibinfo{year}{2022}\natexlab{}.
\newblock \showarticletitle{Interactive Model Cards: A Human-Centered Approach to Model Documentation}.
\newblock \bibinfo{journal}{\emph{arXiv [cs.HC]}} (\bibinfo{date}{May} \bibinfo{year}{2022}).
\newblock


\bibitem[Dabbish et~al\mbox{.}(2012)]%
        {Dabbish2012-ao}
\bibfield{author}{\bibinfo{person}{Laura Dabbish}, \bibinfo{person}{Colleen Stuart}, \bibinfo{person}{Jason Tsay}, {and} \bibinfo{person}{Jim Herbsleb}.} \bibinfo{year}{2012}\natexlab{}.
\newblock \showarticletitle{Social coding in {GitHub}: transparency and collaboration in an open software repository}. In \bibinfo{booktitle}{\emph{Proceedings of the ACM 2012 conference on Computer Supported Cooperative Work}}. \bibinfo{publisher}{ACM}, \bibinfo{address}{New York, NY, USA}.
\newblock


\bibitem[Davani et~al\mbox{.}(2024)]%
        {Davani2024-rr}
\bibfield{author}{\bibinfo{person}{Aida~Mostafazadeh Davani}, \bibinfo{person}{Mark Díaz}, \bibinfo{person}{Dylan Baker}, {and} \bibinfo{person}{Vinodkumar Prabhakaran}.} \bibinfo{year}{2024}\natexlab{}.
\newblock \showarticletitle{{D3CODE}: Disentangling disagreements in data across cultures on offensiveness detection and evaluation}.
\newblock \bibinfo{journal}{\emph{arXiv [cs.CL]}} (\bibinfo{date}{April} \bibinfo{year}{2024}).
\newblock


\bibitem[Deng et~al\mbox{.}(2023)]%
        {Deng2023-wo}
\bibfield{author}{\bibinfo{person}{Wesley~Hanwen Deng}, \bibinfo{person}{Nur Yildirim}, \bibinfo{person}{Monica Chang}, \bibinfo{person}{Motahhare Eslami}, \bibinfo{person}{Kenneth Holstein}, {and} \bibinfo{person}{Michael Madaio}.} \bibinfo{year}{2023}\natexlab{}.
\newblock \showarticletitle{Investigating practices and opportunities for cross-functional collaboration around {AI} fairness in industry practice}. In \bibinfo{booktitle}{\emph{2023 ACM Conference on Fairness, Accountability, and Transparency}}. \bibinfo{publisher}{ACM}, \bibinfo{address}{New York, NY, USA}, \bibinfo{pages}{705--716}.
\newblock


\bibitem[Diaz and Madaio(2024)]%
        {diaz2024scaling}
\bibfield{author}{\bibinfo{person}{Fernando Diaz} {and} \bibinfo{person}{Michael Madaio}.} \bibinfo{year}{2024}\natexlab{}.
\newblock \showarticletitle{Scaling laws do not scale}. In \bibinfo{booktitle}{\emph{Proceedings of the AAAI/ACM Conference on AI, Ethics, and Society}}, Vol.~\bibinfo{volume}{7}. \bibinfo{pages}{341--357}.
\newblock


\bibitem[Face(2024a)]%
        {hugging}
\bibfield{author}{\bibinfo{person}{Hugging Face}.} \bibinfo{year}{2024}\natexlab{a}.
\newblock \bibinfo{title}{Hugging Face Model Cards}.
\newblock
\urldef\tempurl%
\url{https://huggingface.co/docs/hub/en/model-cards}
\showURL{%
\tempurl}


\bibitem[Face(2024b)]%
        {Face2024-op}
\bibfield{author}{\bibinfo{person}{Hugging Face}.} \bibinfo{year}{2024}\natexlab{b}.
\newblock \bibinfo{title}{Reports on the Hub: A First Look at Self-governance in Open Source {AI} Development}.
\newblock \bibinfo{howpublished}{\url{https://huggingface.co/blog/frimelle/self-governance-open-source-ai}}.
\newblock
\newblock
\shownote{Accessed: 2024-12-27}.


\bibitem[Filippova and Cho(2015)]%
        {Filippova2015-bj}
\bibfield{author}{\bibinfo{person}{Anna Filippova} {and} \bibinfo{person}{Hichang Cho}.} \bibinfo{year}{2015}\natexlab{}.
\newblock \showarticletitle{Mudslinging and manners: Unpacking conflict in free and open source software}. In \bibinfo{booktitle}{\emph{Proceedings of the 18th ACM Conference on Computer Supported Cooperative Work \& Social Computing}}. \bibinfo{publisher}{ACM}, \bibinfo{address}{New York, NY, USA}.
\newblock


\bibitem[Frluckaj et~al\mbox{.}(2022)]%
        {Frluckaj2022-gp}
\bibfield{author}{\bibinfo{person}{Hana Frluckaj}, \bibinfo{person}{Laura Dabbish}, \bibinfo{person}{David~Gray Widder}, \bibinfo{person}{Huilian~Sophie Qiu}, {and} \bibinfo{person}{James~D Herbsleb}.} \bibinfo{year}{2022}\natexlab{}.
\newblock \showarticletitle{Gender and participation in open source software development}.
\newblock \bibinfo{journal}{\emph{Proc. ACM Hum. Comput. Interact.}} \bibinfo{volume}{6}, \bibinfo{number}{CSCW2} (\bibinfo{date}{Nov.} \bibinfo{year}{2022}), \bibinfo{pages}{1--31}.
\newblock


\bibitem[Frluckaj et~al\mbox{.}(2024)]%
        {Frluckaj2024-hh}
\bibfield{author}{\bibinfo{person}{Hana Frluckaj}, \bibinfo{person}{Nikki Stevens}, \bibinfo{person}{James Howison}, {and} \bibinfo{person}{Laura Dabbish}.} \bibinfo{year}{2024}\natexlab{}.
\newblock \showarticletitle{Paradoxes of openness: Trans experiences in open source software}.
\newblock \bibinfo{journal}{\emph{Proc. ACM Hum. Comput. Interact.}} \bibinfo{volume}{8}, \bibinfo{number}{CSCW2} (\bibinfo{date}{Nov.} \bibinfo{year}{2024}), \bibinfo{pages}{1--24}.
\newblock


\bibitem[Ganguli et~al\mbox{.}(2022a)]%
        {Ganguli2022-ox}
\bibfield{author}{\bibinfo{person}{Deep Ganguli}, \bibinfo{person}{Danny Hernandez}, \bibinfo{person}{Liane Lovitt}, \bibinfo{person}{Amanda Askell}, \bibinfo{person}{Yuntao Bai}, \bibinfo{person}{Anna Chen}, \bibinfo{person}{Tom Conerly}, \bibinfo{person}{Nova Dassarma}, \bibinfo{person}{Dawn Drain}, \bibinfo{person}{Nelson Elhage}, \bibinfo{person}{Sheer El~Showk}, \bibinfo{person}{Stanislav Fort}, \bibinfo{person}{Zac Hatfield-Dodds}, \bibinfo{person}{Tom Henighan}, \bibinfo{person}{Scott Johnston}, \bibinfo{person}{Andy Jones}, \bibinfo{person}{Nicholas Joseph}, \bibinfo{person}{Jackson Kernian}, \bibinfo{person}{Shauna Kravec}, \bibinfo{person}{Ben Mann}, \bibinfo{person}{Neel Nanda}, \bibinfo{person}{Kamal Ndousse}, \bibinfo{person}{Catherine Olsson}, \bibinfo{person}{Daniela Amodei}, \bibinfo{person}{Tom Brown}, \bibinfo{person}{Jared Kaplan}, \bibinfo{person}{Sam McCandlish}, \bibinfo{person}{Christopher Olah}, \bibinfo{person}{Dario Amodei}, {and} \bibinfo{person}{Jack Clark}.}
  \bibinfo{year}{2022}\natexlab{a}.
\newblock \showarticletitle{Predictability and surprise in large generative models}. In \bibinfo{booktitle}{\emph{2022 ACM Conference on Fairness, Accountability, and Transparency}}. \bibinfo{publisher}{ACM}, \bibinfo{address}{New York, NY, USA}.
\newblock


\bibitem[Ganguli et~al\mbox{.}(2022b)]%
        {Ganguli2022-hp}
\bibfield{author}{\bibinfo{person}{Deep Ganguli}, \bibinfo{person}{Liane Lovitt}, \bibinfo{person}{Jackson Kernion}, \bibinfo{person}{Amanda Askell}, \bibinfo{person}{Yuntao Bai}, \bibinfo{person}{Saurav Kadavath}, \bibinfo{person}{Ben Mann}, \bibinfo{person}{Ethan Perez}, \bibinfo{person}{Nicholas Schiefer}, \bibinfo{person}{Kamal Ndousse}, \bibinfo{person}{Andy Jones}, \bibinfo{person}{Sam Bowman}, \bibinfo{person}{Anna Chen}, \bibinfo{person}{Tom Conerly}, \bibinfo{person}{Nova DasSarma}, \bibinfo{person}{Dawn Drain}, \bibinfo{person}{Nelson Elhage}, \bibinfo{person}{Sheer El-Showk}, \bibinfo{person}{Stanislav Fort}, \bibinfo{person}{Zac Hatfield-Dodds}, \bibinfo{person}{Tom Henighan}, \bibinfo{person}{Danny Hernandez}, \bibinfo{person}{Tristan Hume}, \bibinfo{person}{Josh Jacobson}, \bibinfo{person}{Scott Johnston}, \bibinfo{person}{Shauna Kravec}, \bibinfo{person}{Catherine Olsson}, \bibinfo{person}{Sam Ringer}, \bibinfo{person}{Eli Tran-Johnson}, \bibinfo{person}{Dario Amodei}, \bibinfo{person}{Tom
  Brown}, \bibinfo{person}{Nicholas Joseph}, \bibinfo{person}{Sam McCandlish}, \bibinfo{person}{Chris Olah}, \bibinfo{person}{Jared Kaplan}, {and} \bibinfo{person}{Jack Clark}.} \bibinfo{year}{2022}\natexlab{b}.
\newblock \showarticletitle{Red Teaming Language Models to Reduce Harms: Methods, Scaling Behaviors, and Lessons Learned}.
\newblock \bibinfo{journal}{\emph{arXiv [cs.CL]}} (\bibinfo{date}{Aug.} \bibinfo{year}{2022}).
\newblock


\bibitem[Gebru et~al\mbox{.}(2018)]%
        {Gebru2018-jt}
\bibfield{author}{\bibinfo{person}{Timnit Gebru}, \bibinfo{person}{Jamie Morgenstern}, \bibinfo{person}{Briana Vecchione}, \bibinfo{person}{Jennifer~Wortman Vaughan}, \bibinfo{person}{Hanna Wallach}, \bibinfo{person}{Hal Daumé, III}, {and} \bibinfo{person}{Kate Crawford}.} \bibinfo{year}{2018}\natexlab{}.
\newblock \showarticletitle{Datasheets for Datasets}.
\newblock \bibinfo{journal}{\emph{arXiv [cs.DB]}} (\bibinfo{date}{March} \bibinfo{year}{2018}).
\newblock


\bibitem[Gebru et~al\mbox{.}(2021)]%
        {Gebru2021-oz}
\bibfield{author}{\bibinfo{person}{Timnit Gebru}, \bibinfo{person}{Jamie Morgenstern}, \bibinfo{person}{Briana Vecchione}, \bibinfo{person}{Jennifer~Wortman Vaughan}, \bibinfo{person}{Hanna Wallach}, \bibinfo{person}{Hal~Daumé Iii}, {and} \bibinfo{person}{Kate Crawford}.} \bibinfo{year}{2021}\natexlab{}.
\newblock \showarticletitle{Datasheets for datasets}.
\newblock \bibinfo{journal}{\emph{Commun. ACM}} \bibinfo{volume}{64}, \bibinfo{number}{12} (\bibinfo{date}{Dec.} \bibinfo{year}{2021}), \bibinfo{pages}{86--92}.
\newblock


\bibitem[Geiger et~al\mbox{.}(2018)]%
        {Geiger2018-bg}
\bibfield{author}{\bibinfo{person}{R~Stuart Geiger}, \bibinfo{person}{Nelle Varoquaux}, \bibinfo{person}{Charlotte Mazel-Cabasse}, {and} \bibinfo{person}{Chris Holdgraf}.} \bibinfo{year}{2018}\natexlab{}.
\newblock \showarticletitle{The types, roles, and practices of documentation in data analytics open source software libraries: A collaborative ethnography of documentation work}.
\newblock \bibinfo{journal}{\emph{Comput. Support. Coop. Work}} \bibinfo{volume}{27}, \bibinfo{number}{3-6} (\bibinfo{date}{Dec.} \bibinfo{year}{2018}), \bibinfo{pages}{767--802}.
\newblock


\bibitem[Germonprez et~al\mbox{.}(2018)]%
        {Germonprez2018-pm}
\bibfield{author}{\bibinfo{person}{Matt Germonprez}, \bibinfo{person}{Georg J~P Link}, \bibinfo{person}{Kevin Lumbard}, {and} \bibinfo{person}{S Goggins}.} \bibinfo{year}{2018}\natexlab{}.
\newblock \bibinfo{title}{Eight Observations and 24 Research Questions About Open Source Projects}.
\newblock \bibinfo{numpages}{22}~pages.
\newblock


\bibitem[Gorwa and Veale(2023)]%
        {Gorwa2023-my}
\bibfield{author}{\bibinfo{person}{Robert Gorwa} {and} \bibinfo{person}{Michael Veale}.} \bibinfo{year}{2023}\natexlab{}.
\newblock \showarticletitle{Moderating model marketplaces: Platform governance puzzles for {AI} intermediaries}.
\newblock \bibinfo{journal}{\emph{arXiv [cs.CY]}} (\bibinfo{date}{Nov.} \bibinfo{year}{2023}).
\newblock


\bibitem[Gray~Widder et~al\mbox{.}(2023)]%
        {Gray-Widder2023-xf}
\bibfield{author}{\bibinfo{person}{David Gray~Widder}, \bibinfo{person}{Sarah West}, {and} \bibinfo{person}{Meredith Whittaker}.} \bibinfo{year}{2023}\natexlab{}.
\newblock \showarticletitle{Open (for business): Big tech, concentrated power, and the political economy of open {AI}}.
\newblock \bibinfo{journal}{\emph{SSRN Electron. J.}} (\bibinfo{date}{Aug.} \bibinfo{year}{2023}).
\newblock


\bibitem[Grill(2024)]%
        {Grill2024-wx}
\bibfield{author}{\bibinfo{person}{Gabriel Grill}.} \bibinfo{year}{2024}\natexlab{}.
\newblock \showarticletitle{Constructing capabilities: The politics of testing infrastructures for generative {AI}}. In \bibinfo{booktitle}{\emph{The 2024 ACM Conference on Fairness, Accountability, and Transparency}}. \bibinfo{publisher}{ACM}, \bibinfo{address}{New York, NY, USA}.
\newblock


\bibitem[Guerdan et~al\mbox{.}(2025)]%
        {Guerdan2025-ni}
\bibfield{author}{\bibinfo{person}{Luke Guerdan}, \bibinfo{person}{Solon Barocas}, \bibinfo{person}{Kenneth Holstein}, \bibinfo{person}{Hanna Wallach}, \bibinfo{person}{Zhiwei~Steven Wu}, {and} \bibinfo{person}{Alexandra Chouldechova}.} \bibinfo{year}{2025}\natexlab{}.
\newblock \showarticletitle{Validating {LLM}-as-a-judge systems under rating indeterminacy}.
\newblock \bibinfo{journal}{\emph{arXiv [cs.LG]}} (\bibinfo{date}{Oct.} \bibinfo{year}{2025}).
\newblock


\bibitem[Halfaker et~al\mbox{.}(2009)]%
        {Halfaker2009-hi}
\bibfield{author}{\bibinfo{person}{Aaron Halfaker}, \bibinfo{person}{Aniket Kittur}, \bibinfo{person}{Robert Kraut}, {and} \bibinfo{person}{John Riedl}.} \bibinfo{year}{2009}\natexlab{}.
\newblock \showarticletitle{A jury of your peers: quality, experience and ownership in Wikipedia}. In \bibinfo{booktitle}{\emph{Proceedings of the 5th International Symposium on Wikis and Open Collaboration}}. \bibinfo{publisher}{ACM}, \bibinfo{address}{New York, NY, USA}.
\newblock


\bibitem[Heger et~al\mbox{.}(2022)]%
        {Heger2022-em}
\bibfield{author}{\bibinfo{person}{Amy~K Heger}, \bibinfo{person}{Liz~B Marquis}, \bibinfo{person}{Mihaela Vorvoreanu}, \bibinfo{person}{Hanna Wallach}, {and} \bibinfo{person}{Jennifer~Wortman Vaughan}.} \bibinfo{year}{2022}\natexlab{}.
\newblock \showarticletitle{Understanding machine learning practitioners' data documentation perceptions, needs, challenges, and desiderata}.
\newblock \bibinfo{journal}{\emph{arXiv [cs.HC]}} (\bibinfo{date}{June} \bibinfo{year}{2022}).
\newblock


\bibitem[Hind et~al\mbox{.}(2019)]%
        {Hind2019-jn}
\bibfield{author}{\bibinfo{person}{M Hind}, \bibinfo{person}{Stephanie Houde}, \bibinfo{person}{Jacquelyn Martino}, \bibinfo{person}{A Mojsilovic}, \bibinfo{person}{David Piorkowski}, \bibinfo{person}{John~T Richards}, {and} \bibinfo{person}{Kush~R Varshney}.} \bibinfo{year}{2019}\natexlab{}.
\newblock \bibinfo{title}{Experiences with Improving the Transparency of {AI} Models and Services}.
\newblock


\bibitem[Hind et~al\mbox{.}(2018)]%
        {Hind2018-jc}
\bibfield{author}{\bibinfo{person}{M Hind}, \bibinfo{person}{S Mehta}, \bibinfo{person}{A Mojsilovic}, \bibinfo{person}{R Nair}, \bibinfo{person}{K Ramamurthy}, \bibinfo{person}{Alexandra Olteanu}, {and} \bibinfo{person}{Kush~R Varshney}.} \bibinfo{year}{2018}\natexlab{}.
\newblock \bibinfo{title}{Increasing Trust in {AI} Services through Supplier's Declarations of Conformity}.
\newblock


\bibitem[Hopkins et~al\mbox{.}(2023)]%
        {Hopkins2023-qa}
\bibfield{author}{\bibinfo{person}{Aspen Hopkins}, \bibinfo{person}{Fred Hohman}, \bibinfo{person}{Luca Zappella}, \bibinfo{person}{Xavier~Suau Cuadros}, {and} \bibinfo{person}{Dominik Moritz}.} \bibinfo{year}{2023}\natexlab{}.
\newblock \showarticletitle{Designing data: Proactive data collection and iteration for machine learning}.
\newblock \bibinfo{journal}{\emph{arXiv [cs.HC]}} (\bibinfo{date}{Jan.} \bibinfo{year}{2023}).
\newblock


\bibitem[Howard(2023)]%
        {Howard2023-on}
\bibfield{author}{\bibinfo{person}{Jeremy Howard}.} \bibinfo{year}{2023}\natexlab{}.
\newblock \bibinfo{title}{{AI} Safety and the Age of Dislightenment}.
\newblock \bibinfo{howpublished}{\url{https://www.fast.ai/posts/2023-11-07-dislightenment.html}}.
\newblock
\newblock
\shownote{Accessed: 2025-1-14}.


\bibitem[Huang et~al\mbox{.}(2021)]%
        {Huang2021-ow}
\bibfield{author}{\bibinfo{person}{Yu Huang}, \bibinfo{person}{Denae Ford}, {and} \bibinfo{person}{Thomas Zimmermann}.} \bibinfo{year}{2021}\natexlab{}.
\newblock \showarticletitle{Leaving my fingerprints: Motivations and challenges of contributing to {OSS} for social good}. In \bibinfo{booktitle}{\emph{2021 IEEE/ACM 43rd International Conference on Software Engineering (ICSE)}}. \bibinfo{publisher}{IEEE}, \bibinfo{pages}{1020--1032}.
\newblock


\bibitem[Hutchinson et~al\mbox{.}(2022)]%
        {hutchinson2022evaluation}
\bibfield{author}{\bibinfo{person}{Ben Hutchinson}, \bibinfo{person}{Negar Rostamzadeh}, \bibinfo{person}{Christina Greer}, \bibinfo{person}{Katherine Heller}, {and} \bibinfo{person}{Vinodkumar Prabhakaran}.} \bibinfo{year}{2022}\natexlab{}.
\newblock \showarticletitle{Evaluation gaps in machine learning practice}. In \bibinfo{booktitle}{\emph{Proceedings of the 2022 ACM conference on fairness, accountability, and transparency}}. \bibinfo{pages}{1859--1876}.
\newblock


\bibitem[Intelligence(2024)]%
        {Intelligence2024-io}
\bibfield{author}{\bibinfo{person}{Robust Intelligence}.} \bibinfo{year}{2024}\natexlab{}.
\newblock \bibinfo{title}{Fine-Tuning {LLMs} Breaks Their Safety and Security Alignment — Robust Intelligence}.
\newblock \bibinfo{howpublished}{\url{https://www.robustintelligence.com/blog-posts/fine-tuning-llms-breaks-their-safety-and-security-alignment}}.
\newblock
\newblock
\shownote{Accessed: 2025-11-17}.


\bibitem[Jacobs and Wallach(2021)]%
        {jacobs2021measurement}
\bibfield{author}{\bibinfo{person}{Abigail~Z Jacobs} {and} \bibinfo{person}{Hanna Wallach}.} \bibinfo{year}{2021}\natexlab{}.
\newblock \showarticletitle{Measurement and fairness}. In \bibinfo{booktitle}{\emph{Proceedings of the 2021 ACM conference on fairness, accountability, and transparency}}. \bibinfo{pages}{375--385}.
\newblock


\bibitem[Jahn et~al\mbox{.}(2025)]%
        {Jahn2025-br}
\bibfield{author}{\bibinfo{person}{Leonie Jahn}, \bibinfo{person}{Philip Engelbutzeder}, \bibinfo{person}{Lea~Katharina Michel}, \bibinfo{person}{Sebastian Prost}, \bibinfo{person}{Michael~Bernard Twidale}, \bibinfo{person}{Dave Randall}, {and} \bibinfo{person}{Volker Wulf}.} \bibinfo{year}{2025}\natexlab{}.
\newblock \showarticletitle{Blending Code and Cause: Understanding the Dynamic Motivations of Volunteer Developers in community-driven {FOSS} projects}. In \bibinfo{booktitle}{\emph{Proceedings of the 2025 CHI Conference on Human Factors in Computing Systems}}. \bibinfo{publisher}{ACM}, \bibinfo{address}{New York, NY, USA}, \bibinfo{pages}{1--17}.
\newblock


\bibitem[Jahn et~al\mbox{.}(2024)]%
        {Jahn2024-no}
\bibfield{author}{\bibinfo{person}{Leonie Jahn}, \bibinfo{person}{Philip Engelbutzeder}, \bibinfo{person}{Dave Randall}, \bibinfo{person}{Yannick Bollmann}, \bibinfo{person}{Vasilis Ntouros}, \bibinfo{person}{Lea~Katharina Michel}, {and} \bibinfo{person}{Volker Wulf}.} \bibinfo{year}{2024}\natexlab{}.
\newblock \showarticletitle{In between users and developers: Serendipitous connections and intermediaries in volunteer-driven open-source software development}. In \bibinfo{booktitle}{\emph{Proceedings of the CHI Conference on Human Factors in Computing Systems}}, Vol.~\bibinfo{volume}{52}. \bibinfo{publisher}{ACM}, \bibinfo{address}{New York, NY, USA}, \bibinfo{pages}{1--15}.
\newblock


\bibitem[Jain et~al\mbox{.}(2024)]%
        {Jain2024-df}
\bibfield{author}{\bibinfo{person}{Nitisha Jain}, \bibinfo{person}{Mubashara Akhtar}, \bibinfo{person}{Joan Giner-Miguelez}, \bibinfo{person}{Rajat Shinde}, \bibinfo{person}{Joaquin Vanschoren}, \bibinfo{person}{Steffen Vogler}, \bibinfo{person}{Sujata Goswami}, \bibinfo{person}{Yuhan Rao}, \bibinfo{person}{Tim Santos}, \bibinfo{person}{Luis Oala}, \bibinfo{person}{Michalis Karamousadakis}, \bibinfo{person}{M Maskey}, \bibinfo{person}{Pierre Marcenac}, \bibinfo{person}{Costanza Conforti}, \bibinfo{person}{Michael Kuchnik}, \bibinfo{person}{Lora Aroyo}, \bibinfo{person}{Omar Benjelloun}, {and} \bibinfo{person}{E Simperl}.} \bibinfo{year}{2024}\natexlab{}.
\newblock \bibinfo{title}{A Standardized Machine-readable Dataset Documentation Format for Responsible {AI}}.
\newblock


\bibitem[Jamieson et~al\mbox{.}(2022)]%
        {Jamieson2022-gf}
\bibfield{author}{\bibinfo{person}{Jack Jamieson}, \bibinfo{person}{Eureka Foong}, {and} \bibinfo{person}{Naomi Yamashita}.} \bibinfo{year}{2022}\natexlab{}.
\newblock \showarticletitle{Maintaining values: Navigating diverse perspectives in value-charged discussions in open source development}.
\newblock \bibinfo{journal}{\emph{Proc. ACM Hum. Comput. Interact.}} \bibinfo{volume}{6}, \bibinfo{number}{CSCW2} (\bibinfo{date}{Nov.} \bibinfo{year}{2022}), \bibinfo{pages}{1--28}.
\newblock


\bibitem[Janicki et~al\mbox{.}(2025)]%
        {Janicki2025-nq}
\bibfield{author}{\bibinfo{person}{Sylvia Janicki}, \bibinfo{person}{Shubhangi Gupta}, {and} \bibinfo{person}{Nassim Parvin}.} \bibinfo{year}{2025}\natexlab{}.
\newblock \showarticletitle{Reflexive data walks: Cultivating feminist ethos through place-based inquiry}.
\newblock \bibinfo{journal}{\emph{Proc. ACM Hum. Comput. Interact.}} \bibinfo{volume}{9}, \bibinfo{number}{2} (\bibinfo{date}{May} \bibinfo{year}{2025}), \bibinfo{pages}{1--28}.
\newblock


\bibitem[Jiang et~al\mbox{.}(2023)]%
        {Jiang2023-ev}
\bibfield{author}{\bibinfo{person}{Wenxin Jiang}, \bibinfo{person}{Nicholas Synovic}, \bibinfo{person}{Matt Hyatt}, \bibinfo{person}{Taylor~R Schorlemmer}, \bibinfo{person}{Rohan Sethi}, \bibinfo{person}{Yung-Hsiang Lu}, \bibinfo{person}{George~K Thiruvathukal}, {and} \bibinfo{person}{James~C Davis}.} \bibinfo{year}{2023}\natexlab{}.
\newblock \showarticletitle{An empirical study of pre-trained model reuse in the Hugging Face deep learning model registry}.
\newblock \bibinfo{journal}{\emph{arXiv [cs.SE]}} (\bibinfo{date}{March} \bibinfo{year}{2023}).
\newblock


\bibitem[Jiao et~al\mbox{.}(2024)]%
        {Jiao2024-eg}
\bibfield{author}{\bibinfo{person}{Junfeng Jiao}, \bibinfo{person}{S Afroogh}, \bibinfo{person}{Yiming Xu}, {and} \bibinfo{person}{Connor Phillips}.} \bibinfo{year}{2024}\natexlab{}.
\newblock \bibinfo{title}{Navigating {LLM} Ethics: Advancements, Challenges, and Future Directions}.
\newblock


\bibitem[Jung(2025)]%
        {Jung2025-up}
\bibfield{author}{\bibinfo{person}{Kyu-Hwan Jung}.} \bibinfo{year}{2025}\natexlab{}.
\newblock \showarticletitle{Large language models in medicine: Clinical applications, technical challenges, and ethical considerations}.
\newblock \bibinfo{journal}{\emph{Healthc. Inform. Res.}} \bibinfo{volume}{31}, \bibinfo{number}{2} (\bibinfo{date}{April} \bibinfo{year}{2025}), \bibinfo{pages}{114--124}.
\newblock


\bibitem[Kapoor et~al\mbox{.}(2024)]%
        {Kapoor2024-al}
\bibfield{author}{\bibinfo{person}{Sayash Kapoor}, \bibinfo{person}{Rishi Bommasani}, \bibinfo{person}{Kevin Klyman}, \bibinfo{person}{Shayne Longpre}, \bibinfo{person}{Ashwin Ramaswami}, \bibinfo{person}{Peter Cihon}, \bibinfo{person}{Aspen Hopkins}, \bibinfo{person}{Kevin Bankston}, \bibinfo{person}{Stella Biderman}, \bibinfo{person}{Miranda Bogen}, \bibinfo{person}{Rumman Chowdhury}, \bibinfo{person}{Alex Engler}, \bibinfo{person}{Peter Henderson}, \bibinfo{person}{Yacine Jernite}, \bibinfo{person}{Seth Lazar}, \bibinfo{person}{Stefano Maffulli}, \bibinfo{person}{Alondra Nelson}, \bibinfo{person}{Joelle Pineau}, \bibinfo{person}{Aviya Skowron}, \bibinfo{person}{Dawn Song}, \bibinfo{person}{Victor Storchan}, \bibinfo{person}{Daniel Zhang}, \bibinfo{person}{Daniel~E Ho}, \bibinfo{person}{Percy Liang}, {and} \bibinfo{person}{Arvind Narayanan}.} \bibinfo{year}{2024}\natexlab{}.
\newblock \showarticletitle{On the societal impact of open foundation models}.
\newblock \bibinfo{journal}{\emph{arXiv [cs.CY]}} (\bibinfo{date}{Feb.} \bibinfo{year}{2024}).
\newblock


\bibitem[Karkkainen and Joo(2021)]%
        {karkkainenfairface}
\bibfield{author}{\bibinfo{person}{Kimmo Karkkainen} {and} \bibinfo{person}{Jungseock Joo}.} \bibinfo{year}{2021}\natexlab{}.
\newblock \showarticletitle{FairFace: Face Attribute Dataset for Balanced Race, Gender, and Age for Bias Measurement and Mitigation}. In \bibinfo{booktitle}{\emph{Proceedings of the IEEE/CVF Winter Conference on Applications of Computer Vision}}. \bibinfo{pages}{1548--1558}.
\newblock


\bibitem[Kawakami et~al\mbox{.}(2024)]%
        {Kawakami2024-ip}
\bibfield{author}{\bibinfo{person}{Anna Kawakami}, \bibinfo{person}{Daricia Wilkinson}, {and} \bibinfo{person}{Alexandra Chouldechova}.} \bibinfo{year}{2024}\natexlab{}.
\newblock \bibinfo{title}{Do Responsible {AI} Artifacts Advance Stakeholder Goals? Four Key Barriers Perceived by Legal and Civil Stakeholders}.
\newblock


\bibitem[Kay et~al\mbox{.}(2016)]%
        {Kay2016-hb}
\bibfield{author}{\bibinfo{person}{Matthew Kay}, \bibinfo{person}{Tara Kola}, \bibinfo{person}{Jessica~R Hullman}, {and} \bibinfo{person}{Sean~A Munson}.} \bibinfo{year}{2016}\natexlab{}.
\newblock \showarticletitle{When (ish) is My Bus?: User-centered Visualizations of Uncertainty in Everyday, Mobile Predictive Systems}. In \bibinfo{booktitle}{\emph{Proceedings of the 2016 CHI Conference on Human Factors in Computing Systems}}. \bibinfo{publisher}{ACM}, \bibinfo{address}{New York, NY, USA}.
\newblock


\bibitem[Kazman et~al\mbox{.}(2016)]%
        {Kazman2016-im}
\bibfield{author}{\bibinfo{person}{R Kazman}, \bibinfo{person}{Dennis~R Goldenson}, \bibinfo{person}{I Monarch}, \bibinfo{person}{William~R Nichols}, {and} \bibinfo{person}{G Valetto}.} \bibinfo{year}{2016}\natexlab{}.
\newblock \bibinfo{title}{Evaluating the Effects of Architectural Documentation: A Case Study of a Large Scale Open Source Project}.
\newblock \bibinfo{numpages}{220--260}~pages.
\newblock


\bibitem[Khadpe et~al\mbox{.}(2025)]%
        {Khadpe2025-je}
\bibfield{author}{\bibinfo{person}{Pranav Khadpe}, \bibinfo{person}{Olivia Xu}, \bibinfo{person}{Geoff Kaufman}, {and} \bibinfo{person}{Chinmay Kulkarni}.} \bibinfo{year}{2025}\natexlab{}.
\newblock \showarticletitle{\textit{Hug Reports} : Supporting expression of appreciation between users and contributors of open source software packages}.
\newblock \bibinfo{journal}{\emph{Proc. ACM Hum. Comput. Interact.}} \bibinfo{volume}{9}, \bibinfo{number}{2} (\bibinfo{date}{May} \bibinfo{year}{2025}), \bibinfo{pages}{1--32}.
\newblock


\bibitem[Kittur and Kraut(2008)]%
        {Kittur2008-gw}
\bibfield{author}{\bibinfo{person}{Aniket Kittur} {and} \bibinfo{person}{Robert~E Kraut}.} \bibinfo{year}{2008}\natexlab{}.
\newblock \showarticletitle{Harnessing the wisdom of crowds in wikipedia: quality through coordination}. In \bibinfo{booktitle}{\emph{Proceedings of the 2008 ACM conference on Computer supported cooperative work}}. \bibinfo{publisher}{ACM}, \bibinfo{address}{New York, NY, USA}, \bibinfo{pages}{37--46}.
\newblock


\bibitem[Kolides et~al\mbox{.}(2023)]%
        {Kolides2023-rg}
\bibfield{author}{\bibinfo{person}{Adam Kolides}, \bibinfo{person}{Alyna Nawaz}, \bibinfo{person}{Anshu Rathor}, \bibinfo{person}{Denzel Beeman}, \bibinfo{person}{Muzammil Hashmi}, \bibinfo{person}{Sana Fatima}, \bibinfo{person}{David Berdik}, \bibinfo{person}{Mahmoud Al-Ayyoub}, {and} \bibinfo{person}{Yaser Jararweh}.} \bibinfo{year}{2023}\natexlab{}.
\newblock \showarticletitle{Artificial intelligence foundation and pre-trained models: Fundamentals, applications, opportunities, and social impacts}.
\newblock \bibinfo{journal}{\emph{Simul. Model. Pract. Theory}} \bibinfo{volume}{126}, \bibinfo{number}{102754} (\bibinfo{date}{July} \bibinfo{year}{2023}), \bibinfo{pages}{102754}.
\newblock


\bibitem[Kolt et~al\mbox{.}(2024)]%
        {Kolt2024-zw}
\bibfield{author}{\bibinfo{person}{Noam Kolt}, \bibinfo{person}{Markus Anderljung}, \bibinfo{person}{Joslyn Barnhart}, \bibinfo{person}{Asher Brass}, \bibinfo{person}{K Esvelt}, \bibinfo{person}{Gillian~K Hadfield}, \bibinfo{person}{Lennart Heim}, \bibinfo{person}{Mikel Rodriguez}, \bibinfo{person}{Jonas~B Sandbrink}, {and} \bibinfo{person}{Thomas Woodside}.} \bibinfo{year}{2024}\natexlab{}.
\newblock \showarticletitle{Responsible reporting for frontier {AI} development}.
\newblock \bibinfo{journal}{\emph{AAAI/ACM conference Artificial Intelligence, Ethics, and Society}}  \bibinfo{volume}{abs/2404.02675} (\bibinfo{date}{April} \bibinfo{year}{2024}).
\newblock


\bibitem[Laufer et~al\mbox{.}(2025)]%
        {laufer2025anatomy}
\bibfield{author}{\bibinfo{person}{Benjamin Laufer}, \bibinfo{person}{Hamidah Oderinwale}, {and} \bibinfo{person}{Jon Kleinberg}.} \bibinfo{year}{2025}\natexlab{}.
\newblock \showarticletitle{Anatomy of a Machine Learning Ecosystem: 2 Million Models on Hugging Face}.
\newblock \bibinfo{journal}{\emph{arXiv preprint arXiv:2508.06811}} (\bibinfo{year}{2025}).
\newblock


\bibitem[Lee et~al\mbox{.}(2023)]%
        {Lee2023-bb}
\bibfield{author}{\bibinfo{person}{Katherine Lee}, \bibinfo{person}{A~Feder Cooper}, {and} \bibinfo{person}{James Grimmelmann}.} \bibinfo{year}{2023}\natexlab{}.
\newblock \showarticletitle{Talkin' 'bout {AI} generation: Copyright and the generative-{AI} supply chain}.
\newblock \bibinfo{journal}{\emph{arXiv [cs.CY]}} (\bibinfo{date}{Sept.} \bibinfo{year}{2023}).
\newblock


\bibitem[Li et~al\mbox{.}(2025)]%
        {Li2025-hc}
\bibfield{author}{\bibinfo{person}{Megan Li}, \bibinfo{person}{Wendy Bickersteth}, \bibinfo{person}{Ningjing Tang}, \bibinfo{person}{Jason Hong}, \bibinfo{person}{Lorrie Cranor}, \bibinfo{person}{Hong Shen}, {and} \bibinfo{person}{Hoda Heidari}.} \bibinfo{year}{2025}\natexlab{}.
\newblock \showarticletitle{A closer look at the existing risks of Generative {AI}: Mapping the who, what, and how of real-world incidents}.
\newblock \bibinfo{journal}{\emph{arXiv [cs.CY]}} (\bibinfo{date}{May} \bibinfo{year}{2025}).
\newblock


\bibitem[Liang et~al\mbox{.}(2024)]%
        {Liang2024-hw}
\bibfield{author}{\bibinfo{person}{Weixin Liang}, \bibinfo{person}{Nazneen Rajani}, \bibinfo{person}{Xinyu Yang}, \bibinfo{person}{Ezinwanne Ozoani}, \bibinfo{person}{Eric Wu}, \bibinfo{person}{Yiqun Chen}, \bibinfo{person}{Daniel~Scott Smith}, {and} \bibinfo{person}{James Zou}.} \bibinfo{year}{2024}\natexlab{}.
\newblock \showarticletitle{What's documented in {AI}? Systematic Analysis of {32K} {AI} Model Cards}.
\newblock \bibinfo{journal}{\emph{arXiv [cs.SE]}} (\bibinfo{date}{Feb.} \bibinfo{year}{2024}).
\newblock


\bibitem[Liao et~al\mbox{.}(2023)]%
        {Liao2023-pc}
\bibfield{author}{\bibinfo{person}{Q~Vera Liao}, \bibinfo{person}{Hariharan Subramonyam}, \bibinfo{person}{Jennifer Wang}, {and} \bibinfo{person}{Jennifer Wortman~Vaughan}.} \bibinfo{year}{2023}\natexlab{}.
\newblock \showarticletitle{Designerly Understanding: Information Needs for Model Transparency to Support Design Ideation for {AI}-Powered User Experience}. In \bibinfo{booktitle}{\emph{Proceedings of the 2023 CHI Conference on Human Factors in Computing Systems}} \emph{(\bibinfo{series}{CHI '23}, \bibinfo{number}{Article 9})}. \bibinfo{publisher}{Association for Computing Machinery}, \bibinfo{address}{New York, NY, USA}, \bibinfo{pages}{1--21}.
\newblock


\bibitem[Liao and Vaughan(2023)]%
        {Liao2023-fr}
\bibfield{author}{\bibinfo{person}{Q~Vera Liao} {and} \bibinfo{person}{Jennifer~Wortman Vaughan}.} \bibinfo{year}{2023}\natexlab{}.
\newblock \showarticletitle{{AI} transparency in the age of {LLMs}: A human-centered research roadmap}.
\newblock \bibinfo{journal}{\emph{arXiv [cs.HC]}} (\bibinfo{date}{June} \bibinfo{year}{2023}).
\newblock


\bibitem[Liesenfeld and Dingemanse(2024)]%
        {Liesenfeld2024-gj}
\bibfield{author}{\bibinfo{person}{Andreas Liesenfeld} {and} \bibinfo{person}{Mark Dingemanse}.} \bibinfo{year}{2024}\natexlab{}.
\newblock \showarticletitle{Rethinking open source generative {AI}: open washing and the {EU} {AI} Act}. In \bibinfo{booktitle}{\emph{The 2024 ACM Conference on Fairness, Accountability, and Transparency}}. \bibinfo{publisher}{ACM}, \bibinfo{address}{New York, NY, USA}.
\newblock


\bibitem[Longo and Kelley(2015)]%
        {Longo2015-fa}
\bibfield{author}{\bibinfo{person}{Justin Longo} {and} \bibinfo{person}{Tanya~M Kelley}.} \bibinfo{year}{2015}\natexlab{}.
\newblock \bibinfo{title}{Use of {GitHub} as a platform for open collaboration on text documents}.
\newblock


\bibitem[Madaio et~al\mbox{.}(2024)]%
        {Madaio2024-fe}
\bibfield{author}{\bibinfo{person}{Michael Madaio}, \bibinfo{person}{Shivani Kapania}, \bibinfo{person}{Rida Qadri}, \bibinfo{person}{Ding Wang}, \bibinfo{person}{Andrew Zaldivar}, \bibinfo{person}{Remi Denton}, {and} \bibinfo{person}{Lauren Wilcox}.} \bibinfo{year}{2024}\natexlab{}.
\newblock \showarticletitle{Learning about responsible {AI} on-the-job: Learning pathways, orientations, and aspirations}. In \bibinfo{booktitle}{\emph{The 2024 ACM Conference on Fairness, Accountability, and Transparency}}. \bibinfo{publisher}{ACM}, \bibinfo{address}{New York, NY, USA}.
\newblock


\bibitem[Magooda et~al\mbox{.}(2023)]%
        {Magooda2023-as}
\bibfield{author}{\bibinfo{person}{Ahmed Magooda}, \bibinfo{person}{Alec Helyar}, \bibinfo{person}{Kyle Jackson}, \bibinfo{person}{David Sullivan}, \bibinfo{person}{Chad Atalla}, \bibinfo{person}{Emily Sheng}, \bibinfo{person}{Dan Vann}, \bibinfo{person}{Richard Edgar}, \bibinfo{person}{Hamid Palangi}, \bibinfo{person}{Roman Lutz}, \bibinfo{person}{Hongliang Kong}, \bibinfo{person}{Vincent Yun}, \bibinfo{person}{Eslam Kamal}, \bibinfo{person}{Federico Zarfati}, \bibinfo{person}{Hanna Wallach}, \bibinfo{person}{Sarah Bird}, {and} \bibinfo{person}{Mei Chen}.} \bibinfo{year}{2023}\natexlab{}.
\newblock \showarticletitle{A Framework for Automated Measurement of Responsible {AI} Harms in Generative {AI} Applications}.
\newblock \bibinfo{journal}{\emph{arXiv [cs.CL]}} (\bibinfo{date}{Oct.} \bibinfo{year}{2023}).
\newblock


\bibitem[McDonald et~al\mbox{.}(2019)]%
        {mcdonald2019reliability}
\bibfield{author}{\bibinfo{person}{Nora McDonald}, \bibinfo{person}{Sarita Schoenebeck}, {and} \bibinfo{person}{Andrea Forte}.} \bibinfo{year}{2019}\natexlab{}.
\newblock \showarticletitle{Reliability and inter-rater reliability in qualitative research: Norms and guidelines for CSCW and HCI practice}.
\newblock \bibinfo{journal}{\emph{Proceedings of the ACM on human-computer interaction}} \bibinfo{volume}{3}, \bibinfo{number}{CSCW} (\bibinfo{year}{2019}), \bibinfo{pages}{1--23}.
\newblock


\bibitem[McMillan-Major et~al\mbox{.}(2024)]%
        {McMillan-Major2024-ma}
\bibfield{author}{\bibinfo{person}{Angelina McMillan-Major}, \bibinfo{person}{Emily~M Bender}, {and} \bibinfo{person}{Batya Friedman}.} \bibinfo{year}{2024}\natexlab{}.
\newblock \showarticletitle{Data statements: From technical concept to community practice}.
\newblock \bibinfo{journal}{\emph{ACM J. Responsib. Comput.}} \bibinfo{volume}{1}, \bibinfo{number}{1} (\bibinfo{date}{March} \bibinfo{year}{2024}), \bibinfo{pages}{1--17}.
\newblock


\bibitem[McMillan-Major et~al\mbox{.}(2021)]%
        {McMillan-Major2021-ux}
\bibfield{author}{\bibinfo{person}{Angelina McMillan-Major}, \bibinfo{person}{Salomey Osei}, \bibinfo{person}{Juan~Diego Rodriguez}, \bibinfo{person}{Pawan~Sasanka Ammanamanchi}, \bibinfo{person}{Sebastian Gehrmann}, {and} \bibinfo{person}{Yacine Jernite}.} \bibinfo{year}{2021}\natexlab{}.
\newblock \showarticletitle{Reusable templates and guides for documenting datasets and models for natural language processing and generation: A case study of the {HuggingFace} and {GEM} data and model cards}. In \bibinfo{booktitle}{\emph{Proceedings of the 1st Workshop on Natural Language Generation, Evaluation, and Metrics (GEM 2021)}}. \bibinfo{publisher}{Association for Computational Linguistics}, \bibinfo{address}{Stroudsburg, PA, USA}, \bibinfo{pages}{121--135}.
\newblock


\bibitem[Meta(2024)]%
        {Meta}
\bibfield{author}{\bibinfo{person}{Meta}.} \bibinfo{year}{2024}\natexlab{}.
\newblock \bibinfo{title}{Llama 3}.
\newblock \bibinfo{howpublished}{\url{https://www.llama.com/docs/model-cards-and-prompt-formats/meta-llama-3/}}.
\newblock
\newblock
\shownote{Accessed: 2025-9-9}.


\bibitem[Miceli et~al\mbox{.}(2020)]%
        {miceli2020between}
\bibfield{author}{\bibinfo{person}{Milagros Miceli}, \bibinfo{person}{Martin Schuessler}, {and} \bibinfo{person}{Tianling Yang}.} \bibinfo{year}{2020}\natexlab{}.
\newblock \showarticletitle{Between subjectivity and imposition: Power dynamics in data annotation for computer vision}.
\newblock \bibinfo{journal}{\emph{Proceedings of the ACM on Human-Computer Interaction}} \bibinfo{volume}{4}, \bibinfo{number}{CSCW2} (\bibinfo{year}{2020}), \bibinfo{pages}{1--25}.
\newblock


\bibitem[Mitchell et~al\mbox{.}(2019a)]%
        {Mitchell2019-cv}
\bibfield{author}{\bibinfo{person}{Margaret Mitchell}, \bibinfo{person}{Simone Wu}, \bibinfo{person}{Andrew Zaldivar}, \bibinfo{person}{Parker Barnes}, \bibinfo{person}{Lucy Vasserman}, \bibinfo{person}{Ben Hutchinson}, \bibinfo{person}{Elena Spitzer}, \bibinfo{person}{Inioluwa~Deborah Raji}, {and} \bibinfo{person}{Timnit Gebru}.} \bibinfo{year}{2019}\natexlab{a}.
\newblock \showarticletitle{Model cards for model reporting}.
\newblock  (\bibinfo{year}{2019}).
\newblock


\bibitem[Mitchell et~al\mbox{.}(2019b)]%
        {Mitchell2019-vh}
\bibfield{author}{\bibinfo{person}{Margaret Mitchell}, \bibinfo{person}{Simone Wu}, \bibinfo{person}{Andrew Zaldivar}, \bibinfo{person}{Parker Barnes}, \bibinfo{person}{Lucy Vasserman}, \bibinfo{person}{Ben Hutchinson}, \bibinfo{person}{Elena Spitzer}, \bibinfo{person}{Inioluwa~Deborah Raji}, {and} \bibinfo{person}{Timnit Gebru}.} \bibinfo{year}{2019}\natexlab{b}.
\newblock \showarticletitle{Model cards for model reporting}. In \bibinfo{booktitle}{\emph{Proceedings of the Conference on Fairness, Accountability, and Transparency}}. \bibinfo{publisher}{ACM}, \bibinfo{address}{New York, NY, USA}, \bibinfo{pages}{220--229}.
\newblock


\bibitem[Mohammad(2021)]%
        {Mohammad2021-ns}
\bibfield{author}{\bibinfo{person}{Saif~M Mohammad}.} \bibinfo{year}{2021}\natexlab{}.
\newblock \showarticletitle{Ethics Sheets for {AI} tasks}.
\newblock \bibinfo{journal}{\emph{arXiv [cs.AI]}} (\bibinfo{date}{July} \bibinfo{year}{2021}).
\newblock


\bibitem[Muller et~al\mbox{.}(2021)]%
        {muller2021designing}
\bibfield{author}{\bibinfo{person}{Michael Muller}, \bibinfo{person}{Christine~T Wolf}, \bibinfo{person}{Josh Andres}, \bibinfo{person}{Michael Desmond}, \bibinfo{person}{Narendra~Nath Joshi}, \bibinfo{person}{Zahra Ashktorab}, \bibinfo{person}{Aabhas Sharma}, \bibinfo{person}{Kristina Brimijoin}, \bibinfo{person}{Qian Pan}, \bibinfo{person}{Evelyn Duesterwald}, {et~al\mbox{.}}} \bibinfo{year}{2021}\natexlab{}.
\newblock \showarticletitle{Designing ground truth and the social life of labels}. In \bibinfo{booktitle}{\emph{Proceedings of the 2021 CHI conference on human factors in computing systems}}. \bibinfo{pages}{1--16}.
\newblock


\bibitem[Nist(2024)]%
        {Nist2024-fs}
\bibfield{author}{\bibinfo{person}{Gaithersburg M~D Nist}.} \bibinfo{year}{2024}\natexlab{}.
\newblock \bibinfo{booktitle}{\emph{Artificial intelligence risk management framework: Generative artificial intelligence profile}}.
\newblock \bibinfo{type}{{T}echnical {R}eport}. \bibinfo{address}{Gaithersburg, MD}.
\newblock


\bibitem[Nunes et~al\mbox{.}(2022)]%
        {Nunes2022-th}
\bibfield{author}{\bibinfo{person}{José~Luiz Nunes}, \bibinfo{person}{G~D~J Barbosa}, \bibinfo{person}{Clarisse~Sieckenius de Souza}, \bibinfo{person}{H Lopes}, {and} \bibinfo{person}{Simone Diniz~Junqueira Barbosa}.} \bibinfo{year}{2022}\natexlab{}.
\newblock \bibinfo{title}{Using model cards for ethical reflection: a qualitative exploration}.
\newblock


\bibitem[Nunes et~al\mbox{.}(2024)]%
        {Nunes2024-zf}
\bibfield{author}{\bibinfo{person}{José~Luiz Nunes}, \bibinfo{person}{G~D~J Barbosa}, \bibinfo{person}{C~S~D Souza}, {and} \bibinfo{person}{Simone D~J Barbosa}.} \bibinfo{year}{2024}\natexlab{}.
\newblock \bibinfo{title}{Using Model Cards for ethical reflection on machine learning models: an interview-based study}.
\newblock \bibinfo{numpages}{19}~pages.
\newblock


\bibitem[NVIDIA(2024)]%
        {NVIDIA}
\bibfield{author}{\bibinfo{person}{NVIDIA}.} \bibinfo{year}{2024}\natexlab{}.
\newblock \bibinfo{title}{{StyleGAN3} pretrained models}.
\newblock \bibinfo{howpublished}{\url{https://catalog.ngc.nvidia.com/orgs/nvidia/teams/research/models/stylegan3}}.
\newblock
\newblock
\shownote{Accessed: 2025-9-9}.


\bibitem[Obradovich et~al\mbox{.}(2024)]%
        {Obradovich2024-vq}
\bibfield{author}{\bibinfo{person}{Nick Obradovich}, \bibinfo{person}{Sahib~S Khalsa}, \bibinfo{person}{Waqas Khan}, \bibinfo{person}{Jina Suh}, \bibinfo{person}{Roy~H Perlis}, \bibinfo{person}{Olusola Ajilore}, {and} \bibinfo{person}{Martin~P Paulus}.} \bibinfo{year}{2024}\natexlab{}.
\newblock \showarticletitle{Opportunities and risks of Large Language Models in psychiatry}.
\newblock \bibinfo{journal}{\emph{NPP Digit. Psychiatry Neurosci.}} \bibinfo{volume}{2}, \bibinfo{number}{1} (\bibinfo{date}{May} \bibinfo{year}{2024}), \bibinfo{pages}{1--8}.
\newblock


\bibitem[{Open Source Initiative}(2024)]%
        {Open-Source-Initiative2024-nd}
\bibfield{author}{\bibinfo{person}{{Open Source Initiative}}.} \bibinfo{year}{2024}\natexlab{}.
\newblock \bibinfo{title}{The {MIT} License}.
\newblock \bibinfo{howpublished}{\url{https://opensource.org/license/mit}}.
\newblock
\newblock
\shownote{Accessed: 2025-1-14}.


\bibitem[OpenAI(2024)]%
        {OpenAI}
\bibfield{author}{\bibinfo{person}{OpenAI}.} \bibinfo{year}{2024}\natexlab{}.
\newblock \bibinfo{title}{{GPT}-{4o} System Card}.
\newblock \bibinfo{howpublished}{\url{https://openai.com/index/gpt-4o-system-card/}}.
\newblock
\newblock
\shownote{Accessed: 2025-9-9}.


\bibitem[Osborne et~al\mbox{.}(2024)]%
        {Osborne2024-tf}
\bibfield{author}{\bibinfo{person}{Cailean Osborne}, \bibinfo{person}{Jennifer Ding}, {and} \bibinfo{person}{Hannah~Rose Kirk}.} \bibinfo{year}{2024}\natexlab{}.
\newblock \showarticletitle{The {AI} community building the future? A quantitative analysis of development activity on Hugging Face Hub}.
\newblock \bibinfo{journal}{\emph{J. Comput. Soc. Sci.}} \bibinfo{volume}{7}, \bibinfo{number}{2} (\bibinfo{date}{June} \bibinfo{year}{2024}), \bibinfo{pages}{2067--2105}.
\newblock


\bibitem[Park et~al\mbox{.}(2025)]%
        {Park2025-mz}
\bibfield{author}{\bibinfo{person}{Minjung Park}, \bibinfo{person}{Jodi Forlizzi}, {and} \bibinfo{person}{John Zimmerman}.} \bibinfo{year}{2025}\natexlab{}.
\newblock \showarticletitle{Exploring the innovation opportunities for pre-trained models}. In \bibinfo{booktitle}{\emph{Proceedings of the 2025 ACM Designing Interactive Systems Conference}}. \bibinfo{publisher}{ACM}, \bibinfo{address}{New York, NY, USA}, \bibinfo{pages}{1973--2005}.
\newblock


\bibitem[Patton(2014)]%
        {patton2014qualitative}
\bibfield{author}{\bibinfo{person}{Michael~Quinn Patton}.} \bibinfo{year}{2014}\natexlab{}.
\newblock \bibinfo{booktitle}{\emph{Qualitative research \& evaluation methods: Integrating theory and practice}}.
\newblock \bibinfo{publisher}{Sage publications}.
\newblock


\bibitem[Pawlik et~al\mbox{.}(2015)]%
        {Pawlik2015-qp}
\bibfield{author}{\bibinfo{person}{Aleksandra Pawlik}, \bibinfo{person}{J Segal}, \bibinfo{person}{H Sharp}, {and} \bibinfo{person}{M Petre}.} \bibinfo{year}{2015}\natexlab{}.
\newblock \bibinfo{title}{Crowdsourcing Scientific Software Documentation: A Case Study of the {NumPy} Documentation Project}.
\newblock \bibinfo{numpages}{28--36}~pages.
\newblock


\bibitem[Pinho et~al\mbox{.}(2024)]%
        {Pinho2024-jw}
\bibfield{author}{\bibinfo{person}{Giniele Pinho}, \bibinfo{person}{Aguiar~Jeová Caçula}, \bibinfo{person}{Lucas Costa}, \bibinfo{person}{Igor~Scaliante Wiese}, {and} \bibinfo{person}{Allysson~Allex Araújo}.} \bibinfo{year}{2024}\natexlab{}.
\newblock \bibinfo{title}{Challenges and Solutions of Free and Open Source Software Documentation: A Systematic Mapping Study}.
\newblock \bibinfo{numpages}{114--125}~pages.
\newblock


\bibitem[Pistilli et~al\mbox{.}(2023)]%
        {Pistilli2023-xx}
\bibfield{author}{\bibinfo{person}{Giada Pistilli}, \bibinfo{person}{Carlos Muñoz~Ferrandis}, \bibinfo{person}{Yacine Jernite}, {and} \bibinfo{person}{Margaret Mitchell}.} \bibinfo{year}{2023}\natexlab{}.
\newblock \showarticletitle{Stronger together: On the articulation of ethical charters, legal tools, and technical documentation in {ML}}. In \bibinfo{booktitle}{\emph{2023 ACM Conference on Fairness, Accountability, and Transparency}}. \bibinfo{publisher}{ACM}, \bibinfo{address}{New York, NY, USA}, \bibinfo{pages}{343--354}.
\newblock


\bibitem[Qi et~al\mbox{.}(2024)]%
        {Qi2024-vo}
\bibfield{author}{\bibinfo{person}{Xiangyu Qi}, \bibinfo{person}{Boyi Wei}, \bibinfo{person}{Nicholas Carlini}, \bibinfo{person}{Yangsibo Huang}, \bibinfo{person}{Tinghao Xie}, \bibinfo{person}{Luxi He}, \bibinfo{person}{Matthew Jagielski}, \bibinfo{person}{Milad Nasr}, \bibinfo{person}{Prateek Mittal}, {and} \bibinfo{person}{Peter Henderson}.} \bibinfo{year}{2024}\natexlab{}.
\newblock \showarticletitle{On evaluating the durability of safeguards for open-weight {LLMs}}.
\newblock \bibinfo{journal}{\emph{arXiv [cs.CR]}} (\bibinfo{date}{Dec.} \bibinfo{year}{2024}).
\newblock


\bibitem[Radford et~al\mbox{.}(2019)]%
        {radford2019language}
\bibfield{author}{\bibinfo{person}{Alec Radford}, \bibinfo{person}{Jeffrey Wu}, \bibinfo{person}{Rewon Child}, \bibinfo{person}{David Luan}, \bibinfo{person}{Dario Amodei}, \bibinfo{person}{Ilya Sutskever}, {et~al\mbox{.}}} \bibinfo{year}{2019}\natexlab{}.
\newblock \showarticletitle{Language models are unsupervised multitask learners}.
\newblock \bibinfo{journal}{\emph{OpenAI blog}} \bibinfo{volume}{1}, \bibinfo{number}{8} (\bibinfo{year}{2019}), \bibinfo{pages}{9}.
\newblock


\bibitem[Raji et~al\mbox{.}(2021)]%
        {Raji2021-sd}
\bibfield{author}{\bibinfo{person}{Inioluwa~Deborah Raji}, \bibinfo{person}{Emily~M Bender}, \bibinfo{person}{Amandalynne Paullada}, \bibinfo{person}{Emily Denton}, {and} \bibinfo{person}{Alex Hanna}.} \bibinfo{year}{2021}\natexlab{}.
\newblock \showarticletitle{{AI} and the everything in the whole wide world benchmark}.
\newblock \bibinfo{journal}{\emph{arXiv [cs.LG]}} (\bibinfo{date}{Nov.} \bibinfo{year}{2021}).
\newblock


\bibitem[Roberts et~al\mbox{.}(2006)]%
        {Roberts2006-xr}
\bibfield{author}{\bibinfo{person}{Jeffrey~A Roberts}, \bibinfo{person}{I Hann}, {and} \bibinfo{person}{S Slaughter}.} \bibinfo{year}{2006}\natexlab{}.
\newblock \bibinfo{title}{Understanding the Motivations, Participation, and Performance of Open Source Software Developers: A Longitudinal Study of the Apache Projects}.
\newblock \bibinfo{numpages}{984--999}~pages.
\newblock


\bibitem[Rossi(2004)]%
        {Rossi2004-yk}
\bibfield{author}{\bibinfo{person}{M~A Rossi}.} \bibinfo{year}{2004}\natexlab{}.
\newblock \showarticletitle{Decoding the`` free/open Source ({F}/{OSS}) Software Puzzle'', a Survey of Theoretical and Empirical Contributions}.
\newblock  (\bibinfo{year}{2004}).
\newblock


\bibitem[Seger et~al\mbox{.}(2023)]%
        {Seger2023-ht}
\bibfield{author}{\bibinfo{person}{Elizabeth Seger}, \bibinfo{person}{Aviv Ovadya}, \bibinfo{person}{Ben Garfinkel}, \bibinfo{person}{Divya Siddarth}, {and} \bibinfo{person}{Allan Dafoe}.} \bibinfo{year}{2023}\natexlab{}.
\newblock \showarticletitle{Democratising {AI}: Multiple Meanings, Goals, and Methods}.
\newblock \bibinfo{journal}{\emph{arXiv [cs.AI]}} (\bibinfo{date}{March} \bibinfo{year}{2023}).
\newblock


\bibitem[Smith et~al\mbox{.}(2025)]%
        {smith2025pragmatic}
\bibfield{author}{\bibinfo{person}{Jessie~J Smith}, \bibinfo{person}{Michael Madaio}, \bibinfo{person}{Robin Burke}, {and} \bibinfo{person}{Casey Fiesler}.} \bibinfo{year}{2025}\natexlab{}.
\newblock \showarticletitle{Pragmatic Fairness: Evaluating ML Fairness Within the Constraints of Industry}. In \bibinfo{booktitle}{\emph{Proceedings of the 2025 ACM Conference on Fairness, Accountability, and Transparency}}. \bibinfo{pages}{628--638}.
\newblock


\bibitem[Soden et~al\mbox{.}(2024)]%
        {Soden2024-ma}
\bibfield{author}{\bibinfo{person}{Robert Soden}, \bibinfo{person}{Austin Toombs}, {and} \bibinfo{person}{Michaelanne Thomas}.} \bibinfo{year}{2024}\natexlab{}.
\newblock \showarticletitle{Evaluating interpretive research in {HCI}}.
\newblock \bibinfo{journal}{\emph{Interactions}} \bibinfo{volume}{31}, \bibinfo{number}{1} (\bibinfo{date}{Jan.} \bibinfo{year}{2024}), \bibinfo{pages}{38--42}.
\newblock


\bibitem[Solaiman et~al\mbox{.}(2023)]%
        {Solaiman2023-sa}
\bibfield{author}{\bibinfo{person}{Irene Solaiman}, \bibinfo{person}{Zeerak Talat}, \bibinfo{person}{William Agnew}, \bibinfo{person}{Lama Ahmad}, \bibinfo{person}{Dylan Baker}, \bibinfo{person}{Su~Lin Blodgett}, \bibinfo{person}{Hal Daumé, III}, \bibinfo{person}{Jesse Dodge}, \bibinfo{person}{Ellie Evans}, \bibinfo{person}{Sara Hooker}, \bibinfo{person}{Yacine Jernite}, \bibinfo{person}{Alexandra~Sasha Luccioni}, \bibinfo{person}{Alberto Lusoli}, \bibinfo{person}{Margaret Mitchell}, \bibinfo{person}{Jessica Newman}, \bibinfo{person}{Marie-Therese Png}, \bibinfo{person}{Andrew Strait}, {and} \bibinfo{person}{Apostol Vassilev}.} \bibinfo{year}{2023}\natexlab{}.
\newblock \showarticletitle{Evaluating the Social Impact of Generative {AI} Systems in Systems and Society}.
\newblock \bibinfo{journal}{\emph{arXiv [cs.CY]}} (\bibinfo{date}{June} \bibinfo{year}{2023}).
\newblock


\bibitem[Solyst et~al\mbox{.}(2025)]%
        {solyst2025conduit}
\bibfield{author}{\bibinfo{person}{Jaemarie Solyst}, \bibinfo{person}{Lauren Wilcox}, {and} \bibinfo{person}{Michael Madaio}.} \bibinfo{year}{2025}\natexlab{}.
\newblock \showarticletitle{" The Conduit by which Change Happens": Processes, Barriers, and Support for Interpersonal Learning about Responsible AI}. In \bibinfo{booktitle}{\emph{Proceedings of the 2025 CHI Conference on Human Factors in Computing Systems}}. \bibinfo{pages}{1--15}.
\newblock


\bibitem[Srikumar et~al\mbox{.}(2024)]%
        {Srikumar2024-mh}
\bibfield{author}{\bibinfo{person}{Madhulika Srikumar}, \bibinfo{person}{Jiyoo Chang}, {and} \bibinfo{person}{Kasia Chmielinski}.} \bibinfo{year}{2024}\natexlab{}.
\newblock \bibinfo{title}{Risk Mitigation Strategies for the Open Foundation Model Value Chain}.
\newblock \bibinfo{howpublished}{\url{https://partnershiponai.org/resource/risk-mitigation-strategies-for-the-open-foundation-model-value-chain/}}.
\newblock
\newblock
\shownote{Accessed: 2024-12-19}.


\bibitem[Srnicek(2022)]%
        {Srnicek2022-fp}
\bibfield{author}{\bibinfo{person}{Nick Srnicek}.} \bibinfo{year}{2022}\natexlab{}.
\newblock \showarticletitle{Data, Compute, Labor}.
\newblock In \bibinfo{booktitle}{\emph{Digital Work in the Planetary Market}}. \bibinfo{publisher}{The MIT Press}, \bibinfo{pages}{241--262}.
\newblock


\bibitem[St.~Laurent(2004)]%
        {St-Laurent2004-zx}
\bibfield{author}{\bibinfo{person}{Andrew~M St.~Laurent}.} \bibinfo{year}{2004}\natexlab{}.
\newblock \bibinfo{title}{Understanding Open Source and Free Software Licensing}.
\newblock


\bibitem[Tseng et~al\mbox{.}(2025)]%
        {Tseng2025-ta}
\bibfield{author}{\bibinfo{person}{Emily Tseng}, \bibinfo{person}{Meg Young}, \bibinfo{person}{Marianne~Aubin Le~Quéré}, \bibinfo{person}{Aimee Rinehart}, {and} \bibinfo{person}{Harini Suresh}.} \bibinfo{year}{2025}\natexlab{}.
\newblock \showarticletitle{``ownership, not just happy talk'': Co-designing a participatory large language model for journalism}. In \bibinfo{booktitle}{\emph{Proceedings of the 2025 ACM Conference on Fairness, Accountability, and Transparency}}. \bibinfo{publisher}{ACM}, \bibinfo{address}{New York, NY, USA}, \bibinfo{pages}{3119--3130}.
\newblock


\bibitem[Viegas et~al\mbox{.}(2007)]%
        {Viegas2007-bp}
\bibfield{author}{\bibinfo{person}{Fernanda~B Viegas}, \bibinfo{person}{Martin Wattenberg}, \bibinfo{person}{Jesse Kriss}, {and} \bibinfo{person}{Frank van Ham}.} \bibinfo{year}{2007}\natexlab{}.
\newblock \showarticletitle{Talk before you type: Coordination in Wikipedia}. In \bibinfo{booktitle}{\emph{2007 40th Annual Hawaii International Conference on System Sciences (HICSS'07)}}. \bibinfo{publisher}{IEEE}, \bibinfo{pages}{78--78}.
\newblock


\bibitem[Wallach et~al\mbox{.}(2024)]%
        {Wallach2024-id}
\bibfield{author}{\bibinfo{person}{Hanna Wallach}, \bibinfo{person}{Meera Desai}, \bibinfo{person}{Nicholas Pangakis}, \bibinfo{person}{A~Feder Cooper}, \bibinfo{person}{Angelina Wang}, \bibinfo{person}{Solon Barocas}, \bibinfo{person}{Alexandra Chouldechova}, \bibinfo{person}{Chad Atalla}, \bibinfo{person}{Su~Lin Blodgett}, \bibinfo{person}{Emily Corvi}, \bibinfo{person}{P~Alex Dow}, \bibinfo{person}{Jean Garcia-Gathright}, \bibinfo{person}{Alexandra Olteanu}, \bibinfo{person}{Stefanie Reed}, \bibinfo{person}{Emily Sheng}, \bibinfo{person}{Dan Vann}, \bibinfo{person}{Jennifer~Wortman Vaughan}, \bibinfo{person}{Matthew Vogel}, \bibinfo{person}{Hannah Washington}, {and} \bibinfo{person}{Abigail~Z Jacobs}.} \bibinfo{year}{2024}\natexlab{}.
\newblock \showarticletitle{Evaluating generative {AI} systems is a social science measurement challenge}.
\newblock \bibinfo{journal}{\emph{arXiv [cs.CY]}} (\bibinfo{date}{Nov.} \bibinfo{year}{2024}).
\newblock


\bibitem[Weerts et~al\mbox{.}(2023)]%
        {weerts2023fairlearn}
\bibfield{author}{\bibinfo{person}{Hilde Weerts}, \bibinfo{person}{Miroslav Dud{\'\i}k}, \bibinfo{person}{Richard Edgar}, \bibinfo{person}{Adrin Jalali}, \bibinfo{person}{Roman Lutz}, {and} \bibinfo{person}{Michael Madaio}.} \bibinfo{year}{2023}\natexlab{}.
\newblock \showarticletitle{Fairlearn: Assessing and improving fairness of ai systems}.
\newblock \bibinfo{journal}{\emph{Journal of Machine Learning Research}} \bibinfo{volume}{24}, \bibinfo{number}{257} (\bibinfo{year}{2023}), \bibinfo{pages}{1--8}.
\newblock


\bibitem[Weidinger et~al\mbox{.}(2025)]%
        {Weidinger2025-gs}
\bibfield{author}{\bibinfo{person}{Laura Weidinger}, \bibinfo{person}{Inioluwa~Deborah Raji}, \bibinfo{person}{Hanna Wallach}, \bibinfo{person}{Margaret Mitchell}, \bibinfo{person}{Angelina Wang}, \bibinfo{person}{Olawale Salaudeen}, \bibinfo{person}{Rishi Bommasani}, \bibinfo{person}{Deep Ganguli}, \bibinfo{person}{Sanmi Koyejo}, {and} \bibinfo{person}{William Isaac}.} \bibinfo{year}{2025}\natexlab{}.
\newblock \showarticletitle{Toward an evaluation science for generative {AI} systems}.
\newblock \bibinfo{journal}{\emph{arXiv [cs.AI]}} (\bibinfo{date}{March} \bibinfo{year}{2025}).
\newblock


\bibitem[Weidinger et~al\mbox{.}(2023a)]%
        {Weidinger2023-vz}
\bibfield{author}{\bibinfo{person}{Laura Weidinger}, \bibinfo{person}{Maribeth Rauh}, \bibinfo{person}{Nahema Marchal}, \bibinfo{person}{Arianna Manzini}, \bibinfo{person}{Lisa~Anne Hendricks}, \bibinfo{person}{Juan Mateos-Garcia}, \bibinfo{person}{Stevie Bergman}, \bibinfo{person}{Jackie Kay}, \bibinfo{person}{Conor Griffin}, \bibinfo{person}{Ben Bariach}, \bibinfo{person}{Iason Gabriel}, \bibinfo{person}{Verena Rieser}, {and} \bibinfo{person}{William Isaac}.} \bibinfo{year}{2023}\natexlab{a}.
\newblock \showarticletitle{Sociotechnical Safety Evaluation of Generative {AI} Systems}.
\newblock \bibinfo{journal}{\emph{arXiv [cs.AI]}} (\bibinfo{date}{Oct.} \bibinfo{year}{2023}).
\newblock


\bibitem[Weidinger et~al\mbox{.}(2023b)]%
        {Weidinger2023-od}
\bibfield{author}{\bibinfo{person}{Laura Weidinger}, \bibinfo{person}{Maribeth Rauh}, \bibinfo{person}{Nahema Marchal}, \bibinfo{person}{Arianna Manzini}, \bibinfo{person}{Lisa~Anne Hendricks}, \bibinfo{person}{Juan Mateos-Garcia}, \bibinfo{person}{Stevie Bergman}, \bibinfo{person}{Jackie Kay}, \bibinfo{person}{Conor Griffin}, \bibinfo{person}{Ben Bariach}, \bibinfo{person}{Iason Gabriel}, \bibinfo{person}{Verena Rieser}, {and} \bibinfo{person}{William Isaac}.} \bibinfo{year}{2023}\natexlab{b}.
\newblock \showarticletitle{Sociotechnical Safety Evaluation of Generative {AI} Systems}.
\newblock \bibinfo{journal}{\emph{arXiv [cs.AI]}} (\bibinfo{date}{Oct.} \bibinfo{year}{2023}).
\newblock


\bibitem[Wexler et~al\mbox{.}(2020)]%
        {Wexler2020-op}
\bibfield{author}{\bibinfo{person}{James Wexler}, \bibinfo{person}{Mahima Pushkarna}, \bibinfo{person}{Tolga Bolukbasi}, \bibinfo{person}{Martin Wattenberg}, \bibinfo{person}{Fernanda Viegas}, {and} \bibinfo{person}{Jimbo Wilson}.} \bibinfo{year}{2020}\natexlab{}.
\newblock \showarticletitle{The What-If Tool: Interactive Probing of Machine Learning Models}.
\newblock \bibinfo{journal}{\emph{IEEE Trans. Vis. Comput. Graph.}} \bibinfo{volume}{26}, \bibinfo{number}{1} (\bibinfo{date}{Jan.} \bibinfo{year}{2020}), \bibinfo{pages}{56--65}.
\newblock


\bibitem[Widder and Nafus(2023)]%
        {Widder2023-vb}
\bibfield{author}{\bibinfo{person}{David~Gray Widder} {and} \bibinfo{person}{Dawn Nafus}.} \bibinfo{year}{2023}\natexlab{}.
\newblock \showarticletitle{Dislocated accountabilities in the “AI supply chain”: Modularity and developers’ notions of responsibility}.
\newblock \bibinfo{journal}{\emph{Big Data Soc.}} \bibinfo{volume}{10}, \bibinfo{number}{1} (\bibinfo{date}{Jan.} \bibinfo{year}{2023}).
\newblock


\bibitem[Widder et~al\mbox{.}(2022)]%
        {Widder2022-dg}
\bibfield{author}{\bibinfo{person}{David~Gray Widder}, \bibinfo{person}{Dawn Nafus}, \bibinfo{person}{Laura Dabbish}, {and} \bibinfo{person}{James Herbsleb}.} \bibinfo{year}{2022}\natexlab{}.
\newblock \showarticletitle{Limits and Possibilities for “Ethical {AI”} in Open Source: A Study of Deepfakes}. In \bibinfo{booktitle}{\emph{Proceedings of the 2022 ACM Conference on Fairness, Accountability, and Transparency}} \emph{(\bibinfo{series}{FAccT '22})}. \bibinfo{publisher}{Association for Computing Machinery}, \bibinfo{address}{New York, NY, USA}, \bibinfo{pages}{2035--2046}.
\newblock


\bibitem[Widder et~al\mbox{.}(2024)]%
        {Widder2024-wk}
\bibfield{author}{\bibinfo{person}{David~Gray Widder}, \bibinfo{person}{Meredith Whittaker}, {and} \bibinfo{person}{Sarah~Myers West}.} \bibinfo{year}{2024}\natexlab{}.
\newblock \showarticletitle{Why 'open' {AI} systems are actually closed, and why this matters}.
\newblock \bibinfo{journal}{\emph{Nature}} \bibinfo{volume}{635}, \bibinfo{number}{8040} (\bibinfo{date}{Nov.} \bibinfo{year}{2024}), \bibinfo{pages}{827--833}.
\newblock


\bibitem[Winecoff and Bogen(2025)]%
        {Winecoff2025-yd}
\bibfield{author}{\bibinfo{person}{Amy Winecoff} {and} \bibinfo{person}{Miranda Bogen}.} \bibinfo{year}{2025}\natexlab{}.
\newblock \showarticletitle{Improving governance outcomes through {AI} documentation: Bridging theory and practice}. In \bibinfo{booktitle}{\emph{Proceedings of the 2025 CHI Conference on Human Factors in Computing Systems}}. \bibinfo{publisher}{ACM}, \bibinfo{address}{New York, NY, USA}, \bibinfo{pages}{1--18}.
\newblock


\bibitem[Wong(2025)]%
        {Wong2025-yy}
\bibfield{author}{\bibinfo{person}{Richmond~Y Wong}.} \bibinfo{year}{2025}\natexlab{}.
\newblock \showarticletitle{Towards creating infrastructures for values and ethics work in the production of software technologies}.
\newblock \bibinfo{journal}{\emph{arXiv [cs.HC]}} (\bibinfo{date}{July} \bibinfo{year}{2025}).
\newblock


\bibitem[Xiao et~al\mbox{.}(2024)]%
        {Xiao2024-mr}
\bibfield{author}{\bibinfo{person}{Ziang Xiao}, \bibinfo{person}{Wesley~Hanwen Deng}, \bibinfo{person}{Michelle~S Lam}, \bibinfo{person}{Motahhare Eslami}, \bibinfo{person}{Juho Kim}, \bibinfo{person}{Mina Lee}, {and} \bibinfo{person}{Q~Vera Liao}.} \bibinfo{year}{2024}\natexlab{}.
\newblock \showarticletitle{Human-centered evaluation and auditing of language models}. In \bibinfo{booktitle}{\emph{Extended Abstracts of the CHI Conference on Human Factors in Computing Systems}}. \bibinfo{publisher}{ACM}, \bibinfo{address}{New York, NY, USA}, \bibinfo{pages}{1--6}.
\newblock


\bibitem[Yang et~al\mbox{.}(2024)]%
        {Yang2024-kd}
\bibfield{author}{\bibinfo{person}{Xinyu Yang}, \bibinfo{person}{Weixin Liang}, {and} \bibinfo{person}{James Zou}.} \bibinfo{year}{2024}\natexlab{}.
\newblock \showarticletitle{Navigating Dataset Documentations in {AI}: A Large-Scale Analysis of Dataset Cards on Hugging Face}.
\newblock \bibinfo{journal}{\emph{arXiv [cs.LG]}} (\bibinfo{date}{Jan.} \bibinfo{year}{2024}).
\newblock


\bibitem[Yue et~al\mbox{.}(2025)]%
        {Yue2025-tv}
\bibfield{author}{\bibinfo{person}{Huang Yue}, \bibinfo{person}{Gao Chujie}, \bibinfo{person}{Wu Siyuan}, \bibinfo{person}{Wang Haoran}, \bibinfo{person}{Wang Xiangqi}, \bibinfo{person}{Zhou Yujun}, \bibinfo{person}{Wang Yanbo}, \bibinfo{person}{Ye Jiayi}, \bibinfo{person}{Shi Jiawen}, \bibinfo{person}{Zhang Qihui}, \bibinfo{person}{Li Yuan}, \bibinfo{person}{Bao Han}, \bibinfo{person}{Liu Zhaoyi}, \bibinfo{person}{Guan Tianrui}, \bibinfo{person}{Chen Dongping}, \bibinfo{person}{Chen Ruoxi}, \bibinfo{person}{Guo Kehan}, \bibinfo{person}{Zou Andy}, \bibinfo{person}{Bryan~Hooi Kuen-Yew}, \bibinfo{person}{Xiong Caiming}, \bibinfo{person}{Stengel-Eskin Elias}, \bibinfo{person}{Zhang Hongyang}, \bibinfo{person}{Yin Hongzhi}, \bibinfo{person}{Zhang Huan}, \bibinfo{person}{Yao Huaxiu}, \bibinfo{person}{Yoon Jaehong}, \bibinfo{person}{Zhang Jieyu}, \bibinfo{person}{Shu Kai}, \bibinfo{person}{Zhu Kaijie}, \bibinfo{person}{Krishna Ranjay}, \bibinfo{person}{Swayamdipta Swabha}, \bibinfo{person}{Shi Taiwei},
  \bibinfo{person}{Shi Weijia}, \bibinfo{person}{Li Xiang}, \bibinfo{person}{Li Yiwei}, \bibinfo{person}{Hao Yuexing}, \bibinfo{person}{Hao Yuexing}, \bibinfo{person}{Jia Zhihao}, \bibinfo{person}{Li Zhize}, \bibinfo{person}{Chen Xiuying}, \bibinfo{person}{Tu Zhengzhong}, \bibinfo{person}{Hu Xiyang}, \bibinfo{person}{Zhou Tianyi}, \bibinfo{person}{Zhao Jieyu}, \bibinfo{person}{Sun Lichao}, \bibinfo{person}{Huang Furong}, \bibinfo{person}{Or~Cohen Sasson}, \bibinfo{person}{Sattigeri Prasanna}, \bibinfo{person}{Reuel Anka}, \bibinfo{person}{Lamparth Max}, \bibinfo{person}{Zhao Yue}, \bibinfo{person}{Dziri Nouha}, \bibinfo{person}{Su Yu}, \bibinfo{person}{Sun Huan}, \bibinfo{person}{Ji Heng}, \bibinfo{person}{Xiao Chaowei}, \bibinfo{person}{Bansal Mohit}, \bibinfo{person}{V~Chawla Nitesh}, \bibinfo{person}{Pei Jian}, \bibinfo{person}{Gao Jianfeng}, \bibinfo{person}{Backes Michael}, \bibinfo{person}{S~Yu Philip}, \bibinfo{person}{Neil~Zhenqiang Gong}, \bibinfo{person}{Chen Pin-Yu}, \bibinfo{person}{Li Bo}, {and}
  \bibinfo{person}{Zhang Xiangliang}.} \bibinfo{year}{2025}\natexlab{}.
\newblock \showarticletitle{On the trustworthiness of generative Foundation Models: Guideline, assessment, and perspective}.
\newblock \bibinfo{journal}{\emph{arXiv [cs.CY]}} (\bibinfo{date}{Feb.} \bibinfo{year}{2025}).
\newblock


\bibitem[Zhang et~al\mbox{.}(2020)]%
        {Zhang2020-bs}
\bibfield{author}{\bibinfo{person}{Amy~X Zhang}, \bibinfo{person}{Michael Muller}, {and} \bibinfo{person}{Dakuo Wang}.} \bibinfo{year}{2020}\natexlab{}.
\newblock \showarticletitle{How do data science workers collaborate? Roles, workflows, and tools}.
\newblock \bibinfo{journal}{\emph{Proc. ACM Hum. Comput. Interact.}} \bibinfo{volume}{4}, \bibinfo{number}{CSCW1} (\bibinfo{date}{May} \bibinfo{year}{2020}), \bibinfo{pages}{1--23}.
\newblock


\bibitem[Zhou et~al\mbox{.}(2023)]%
        {Zhou2023-gv}
\bibfield{author}{\bibinfo{person}{Jeffrey Zhou}, \bibinfo{person}{Tianjian Lu}, \bibinfo{person}{Swaroop Mishra}, \bibinfo{person}{Siddhartha Brahma}, \bibinfo{person}{Sujoy Basu}, \bibinfo{person}{Yi Luan}, \bibinfo{person}{Denny Zhou}, {and} \bibinfo{person}{Le Hou}.} \bibinfo{year}{2023}\natexlab{}.
\newblock \showarticletitle{Instruction-following evaluation for Large Language Models}.
\newblock \bibinfo{journal}{\emph{arXiv [cs.CL]}} (\bibinfo{date}{Nov.} \bibinfo{year}{2023}).
\newblock


\end{thebibliography}
\appendix 
\label{Appendix}
\newpage
\section{High-level Themes}
\begin{table}[!htbp]
\centering
\begin{tabular}{@{}p{0.4\linewidth}p{0.5\linewidth}@{}}
\toprule
\textbf{Third-Level Themes} & \textbf{Second-Level Themes} \\
\midrule
Normative and Epistemic Uncertainties in Determining the Documentation Content & Uncertainty in Moving Beyond Established Open-Source Documentation Practices \\
& Uncertainties around What Model Characteristics are Feasible to Measure and Report Faithfully \\
& Uncertainties around Appropriate Levels of Disclosure Amid Conflicting Priorities \\
\midrule
Methodological Uncertainties in Evaluating and Communicating Model Properties & Uncertainties in How to Report Model Capabilities \\
& Uncertainties in How to Report Intended and Unintended Model Use Cases \\
& Uncertainties in How to Report Model Risks, Biases, and Limitations \\
\midrule
Ecosystemic Uncertainties in Allocating Documentation Responsibility & Uncertainties in Who Should Report Bias \\
& Uncertainties in Who Should Report Model Performance Across Different Deployment Contexts \\
& Uncertainties in Who Should be Accountable for Documenting Appropriate Use \\
\bottomrule
\end{tabular}
\caption{Nine second-level themes and three third-level themes we identified through thematic analysis}
\label{tab:themes}
\end{table}

\end{document}